\documentclass[reprint, superscriptaddress, amsmath, amssymb, notitlepage, 10pt, aps]{revtex4-1}

\usepackage{xcolor}
\usepackage{graphicx}
\usepackage{bmpsize}
\usepackage{dcolumn}
\usepackage[T1]{fontenc}
\usepackage{bm}
\usepackage{hyperref}
\usepackage{cleveref}
\usepackage{braket}
\usepackage{mathtools}

\usepackage[normalem]{ulem}

\newcommand{\MOcolor}{teal}

\usepackage[most]{tcolorbox}
\usetikzlibrary{patterns}
\pgfdeclarepatternformonly{mystrikeout}{\pgfqpoint{-1pt}{-1pt}}{\pgfqpoint{11pt}{11pt}}{\pgfqpoint{10pt}{10pt}}%
{
  \pgfsetlinewidth{0.4pt}
  \pgfpathmoveto{\pgfqpoint{0pt}{0pt}}
  \pgfpathlineto{\pgfqpoint{10.1pt}{10.1pt}}
  \pgfusepath{stroke}
}
\newtcolorbox{tcbstrikeout}{breakable,
 enhanced jigsaw,
 opacityback=0,
 parbox=false,
 boxrule=0mm,
 top=0mm,bottom=0pt,left=0pt,right=0pt,
 boxsep=0pt,
 frame hidden,
 finish={\fill[pattern=mystrikeout] (frame.north west) rectangle (frame.south east);},
 coltext=\MOcolor
}

\begin{document}


\title{Drude weights in one-dimensional systems with a single defect}

\author{Kazuaki Takasan}
\affiliation{
Department of Physics, University of California, Berkeley, California 94720, USA}
\affiliation{
Materials Sciences Division, Lawrence Berkeley National Laboratory, Berkeley, CA 94720, USA
}
\affiliation{%
Department of Physics, University of Tokyo, 7-3-1 Hongo, Tokyo 113-0033, Japan}%

\author{Masaki Oshikawa}
\affiliation{
Institute for Solid State Physics, University of Tokyo, Kashiwa, Chiba 277-8581, Japan}
\affiliation{
Kavli Institute for the Physics and Mathematics of the Universe (WPI), University of Tokyo, Kashiwa, Chiba 277-8583, Japan}

\author{Haruki Watanabe}
\email{hwatanabe@g.ecc.u-tokyo.ac.jp}
\affiliation{
Department of Applied Physics, University of Tokyo, Tokyo 113-8656, Japan}

\date{\today}

\begin{abstract}
Ballistic transport of a quantum system can be characterized by Drude weight, which quantifies the response of the system to a uniform electric field in the infinitely long timescale. The Drude weight is often discussed in terms of the Kohn formula, which gives the Drude weight by the derivative of the energy eigenvalue of a finite-size system with the periodic boundary condition in terms of the Aharonov-Bohm flux. Recently, the Kohn formula is generalized to nonlinear responses. However, the nonlinear Drude weight determined by the Kohn formula often diverges in the thermodynamic limit. In order to elucidate the issue, in this work we examine a simple example of a one-dimensional tight-binding model in the presence of a single defect at zero temperature. We find that its linear and non-linear Drude weights given by the Kohn formula (i) depend on the Aharonov-Bohm flux and  (ii) diverge proportionally to a power of the system size. 
We argue that the problem can be attributed to different order of limits. The Drude weight according to the Kohn formula (``Kohn--Drude weight'') indicates the response of a finite-size system to an adiabatic insertion of the Aharonov-Bohm flux. While it is a well-defined physical quantity for a finite-size system, its thermodynamic limit does not always describe the ballistic transport of the bulk. The latter should be rather characterized by a ``bulk Drude weight'' defined by taking the thermodynamic limit first before the zero-frequency limit. While the potential issue of the order of limits has been sometimes discussed within the linear response, the discrepancy between the two limits is amplified in nonlinear Drude weights. 
We demonstrate the importance of the low-energy excitations of $O(1/L)$, which are excluded from the Kohn--Drude weight, in regularizing the bulk Drude weight.
\end{abstract}

\maketitle

\section{Introduction}
Transport phenomenon is one of the most fundamental problems in condensed matter and statistical physics. Compared to the widely-studied linear response, nonlinear ones have been less explored and still are an intriguing topic. For instance, rectification current~\cite{Tan2016, Tokura2018} and high-harmonic generation~\cite{Kruchinin2018, Ghimire2019} originating from nonlinear responses are extensively studied recently. Although the theoretical sides have also been actively investigated and various interesting phenomena have been proposed~\cite{Sodemann2015, Morimoto2016, deJuan2017, Dan2019, Isobe2020, Ahn2020, Takasan2020}, we still do not reach a systematic understanding of them. In particular, the general aspects of the nonlinear responses in quantum many-body systems have been less studied~\cite{Shimizu1}.

The current induced by the uniform electric field is an important subject in the charge transport.
Since electromagnetic waves used to probe materials usually have a wavelength much longer than the
microscopic lengthscale, optical absorption can be related to the conductivity in the uniform (zero wavenumber)
limit at the frequency of the wave. Thus the corresponding conductivity as a function of the frequency is
often called optical conductivity.
In a finite system with the periodic boundary condition, the problem can be formulated as the response towards an insertion of a U(1) Aharonov-Bohm magnetic flux (AB flux). 
In particular, the celebrated Kohn formula~\cite{Kohn1964} relates the response to an adiabatic flux insertion to
the Drude weight.
The Drude weight is the coefficient of the delta function peak at zero frequency in the optical conductivity,
and characterizes the ballistic transport~\cite{Resta_2018}
of the system within the linear response theory.
The Kohn formula and Drude weight have been studied in numerous papers over nearly half a century~\cite{ShastrySutherland-twisted, Sutherland1990, Korepin1991, Castella1995, Narozhny1998, Zotos1999, Bertini2021}.

Recently, nonlinear Drude weights were introduced as a direct generalization of the Drude weight of linear response to higher-order responses~\cite{PhysRevB.102.165137}, and their nature has been further investigated~\cite{WatanabeLiuOshikawa, Tanikawa2021,Tanikawa2021fine,Fava, Liu2021,Urichuk2022,Resta2022}. For noninteracting band insulators, the Drude weights underlie the phenomenon so-called Bloch oscillation~\cite{PhysRevB.102.165137}.
The $N$-th order Drude weight $\tilde{\mathcal{D}}_{(N)}^{L,\theta}$ is the coefficient of the most singular term $\prod_{\ell=1}^N\delta(\omega_\ell)$ in the $N$-th order optical conductivity $\sigma_{(N)}^{L,\theta}(\omega_1,\dots,\omega_N)$  for a fixed $L$, which $\tilde{\mathcal{D}}_{(N)}^{L,\theta}$ can be computed by the generalized Kohn formula $\tilde{\mathcal{D}}_{(N)}^{L,\theta}= L^Nd^{N+1}E_0^{L,\theta}/d\theta^{N+1}$. 
In the studies of the nonlinear responses in the spin-1/2 XXZ Heisenberg chain~\cite{PhysRevB.102.165137,Tanikawa2021,Tanikawa2021fine,Fava,Liu2021,Urichuk2022}, it turns out that its nonlinear Drude weights diverge in the limit of large system size, depending on the order of the response and the value of the anisotropy parameter~\cite{PhysRevB.102.165137,Tanikawa2021,Tanikawa2021fine}. 
While the origin of the divergence in the XXZ chain was discussed in Ref.~\cite{Tanikawa2021}, the general condition, such as whether the many-body interaction is needed or not for the divergence, is still unknown. It is also unclear what happens under a static electric field in the system with the divergent Drude weights. It is important to clarify the above problems in order to reach a systematic understanding of the nonlinear responses.

In order to clarify the issues, in this work we study the (linear and nonlinear) Drude weights in a very
simple setup: single-band tight-binding model with a single defect at zero temperature.
The exact solvability of the model helps understanding of the problem in detail.
Even in such a simple setup, we find the divergence of nonlinear Drude weights in the thermodynamic limit.
We relate the apparent pathological behavior of the nonlinear Drude weights to the
order of the two limits: the zero-frequency limit and the thermodynamic limit.
In order to clarify the difference, we call the Drude weight determined by the Kohn formula,
which characterizes the adiabatic transport of the finite-size system, as \emph{Kohn--Drude weight}.
The thermodynamic limit of the Kohn--Drude weight does not necessarily characterize the ballistic transport
of the bulk, which is quantified by the \emph{bulk Drude weight}.
Roughly speaking, the bulk Drude weight is defined by taking the thermodynamic limit first, before taking the
adiabatic (infinitely slow AB flux insertion) limit.
The bulk Drude weight is generally different from the thermodynamic limit of the Kohn--Drude weight, although
they can be identical in some cases. 
While the importance of the distinction between the two limits has been discussed for
the linear Drude weight~\cite{Oshikawa-Drude,IlievskiProsen-DrudeWeights_CMP2013,Sirker-LesHouches2018},
the discrepancy is amplified in the nonlinear Drude weights, most notably in the divergence of the nonlinear
Kohn--Drude weights in the thermodynamic limit~\cite{PhysRevB.102.165137,Tanikawa2021,Tanikawa2021fine}.
Thanks to the simple exactly solvable setup, we demonstrate explicitly how the bulk Drude weight free from
the pathological features is recovered in
the appropriate thermodynamic limit.

First, we examine the dependence of the ground state energy to the twisted boundary condition in general 1D systems and clarify the relation to the adiabatic transport including the nonlinear responses through the Kohn formula.
We then study the specific case of the single-band tight-binding model with a single defect, in which
the nonlinear Kohn--Drude weights diverge in the thermodynamic limit.
This demonstrates that the many-body interactions are not essential for
the divergence of the nonlinear Kohn--Drude weights. 
Regardless of the details of the defect, the divergence of nonlinear Kohn--Drude weights is much stronger than those observed in the XXZ model. Furthermore, we find that even the linear Kohn--Drude weight shows a pathological behavior, depending strongly on AB flux in the thermodynamic limit.

We also elucidate the physical consequence of the diverging behavior of the Kohn--Drude weights. One might think that the divergence implies an arbitrarily large current response.
To clarify this point, we employ the numerical real-time simulation to study the time evolution of the current under the static electric field.
We find that the induced current is rather suppressed compared to the defect-free case.
The real-time simulations would be the most direct approach to the question of the current response to the uniform electric field.
However, the real-time numerical study of the Drude weight
in the gapless phase of the XXZ model is challenging
because the interaction makes the long-time simulation difficult. In contrast,
the noninteracting nature of our model makes it possible to study the crossover between the adiabatic (Kohn formula) limit
and the bulk limit.
 
For a fixed system size $L$, the Kohn--Drude weight corresponds to the coefficient of the delta function at zero
frequency in the optical conductivity. However, the optical conductivity also contains low-frequency peaks
originating from low-energy excitations with the excitation energy of the order $O(1/L)$.
They can contribute to the bulk Drude weight, since they merge to the zero-frequency delta function when
the thermodynamic limit $L \to \infty$ is taken first.
For the single-defect model, 
we demonstrate that, in the bulk Drude weight thus obtained, the pathological features of the Kohn--Drude weight
such as the dependence on the AB flux and the divergence of the nonlinear Drude weights disappear. 
Therefore the bulk Drude weight indeed behaves as a well-defined bulk quantity.
We also discuss implications of these findings for the pathological behaviors observed in many-body interacting systems such as the XXZ spin chain~\cite{PhysRevB.102.165137,Tanikawa2021,Fava,Tanikawa2021fine}.

This article is organized as follows.
In Sec.~\ref{sec:review}, we review response theory to set up notations. 
In Sec.~\ref{sec:twist}, we examine the dependence of the ground state energy on the twisted boundary condition, and summarize its relation to the adiabatic transport.
We derive general upper bounds of the adiabatic current density and the linear Drude weights in terms of the frequency sum of the optical conductivity.
In Sec.~\ref{sec:model}, we introduce our central model in this study, a single-band tight-binding model with a defect, and analytically show that the model has pathological properties including the divergence of the Kohn--Drude weights. 
In Sec.~\ref{sec:realtime}, we study the physical implication of the divergent Kohn--Drude weights using the numerical real-time simulation. 
In Sec.~\ref{sec:Dbulk}, we
demonstrate how the low-frequency components of the optical conductivity in finite-size systems
contribute to the bulk Drude weight and cancel the pathological behavior of the Kohn--Drude weight.
After discussing the relation between the Kohn--Drude and bulk Drude weights in Sec.~\ref{sec:Discussion},
we summarize this paper in Sec.~\ref{sec:Summary}.  

\section{Review of response theory}
\label{sec:review}
Let us start by summarizing the setting of the systems considered in this work and briefly reviewing the linear and the second-order response theory to set the basis of discussions in the subsequent sections.
\subsection{Setting and definitions}
\label{sec:defs}
Let us consider 1D systems described by a local Hamiltonian $\hat{H}^{L,\theta}$ that contains only finite-ranged hoppings and finite-ranged interactions. We impose the periodic boundary condition (PBC) with the system size $L$. Assuming a global U(1) symmetry for the particle number conservation, we introduce a uniform vector potential $A=\theta/L$.  The $n$-th eigenstate of $\hat{H}^{L,\theta}$ is denoted by $\ket{n^{L,\theta}}$ and its energy eigenvalue is written as $E_n^{L,\theta}$. The ground state is assumed to be unique and corresponds to $n=0$. In the following, the expectation value $\langle\,{\cdot}\,\rangle$  is taken using the the ground state $\ket{0^{L,\theta}}$ of $\hat{H}^{L,\theta}$.

The averaged current operator $\hat{j}^{L,\theta}$ and the kinetic energy operator $\hat{k}^{L,\theta}$ are given by
\begin{align}
&\hat{j}^{L,\theta}\equiv\frac{1}{L}\frac{d\hat{H}^{L,\theta}}{dA}=\frac{d\hat{H}^{L,\theta}}{d\theta},\label{eqj}\\
&\hat{k}^{L,\theta}\equiv\frac{1}{L}\frac{d^2\hat{H}^{L,\theta}}{dA^2}=L\frac{d^2\hat{H}^{L,\theta}}{d\theta^2}\label{eqk}.
\end{align}
We also define $\hat{t}^{L,\theta}$ by the third derivative
\begin{align}
\hat{t}^{L,\theta}\equiv\frac{1}{L}\frac{d^3\hat{H}^{L,\theta}}{dA^3}=L^2\frac{d^3\hat{H}^{L,\theta}}{d\theta^3}\label{eqt}
\end{align}
for a later purpose.

Suppose that a time-dependent electric field $E(t)$ is applied to the system.
In the Fourier space, the current induced by the electric field can be expressed in terms of the optical conductivities $\sigma_{(N)}^{L,\theta}(\omega_1,\dots,\omega_N)$ as 
\begin{align}
j^{L,\theta}(\omega)=&\sum_{N=1}^\infty\frac{2\pi}{N!}\int \frac{d\omega_1}{2\pi}\dots \int \frac{d\omega_N}{2\pi}\delta\Big(\omega-\sum_{\ell=1}^N\omega_\ell\Big)\notag\\
&\quad\times\sigma_{(N)}^{L,\theta}(\omega_1,\dots,\omega_N)\prod_{\ell=1}^NE(\omega_\ell).
\end{align}
Here $N$ represents the order of the response: $N=1$ is the linear response and $N\geq 2$ is a higher-order response. 
We define the $N$-th order Kohn--Drude weight $\tilde{\mathcal{D}}_{(N)}^{L,\theta}$ as the coefficient of the $\pi^N\prod_{\ell=1}^N\delta(\omega_\ell)$ term in $\sigma_{(N)}^{L,\theta}(\omega_1,\dots,\omega_N)$~\cite{PhysRevB.102.165137}.  The Kohn--Drude weight can be computed by the generalized Kohn formula~\cite{Kohn1964,PhysRevB.102.165137}:
\begin{align}
\tilde{\mathcal{D}}_{(N)}^{L,\theta}= \frac{1}{L}\frac{d^{N+1}}{dA^{N+1}}E_0^{L,\theta}=L^N\frac{d^{N+1}}{d\theta^{N+1}}E_0^{L,\theta}.
\label{D_N}
\end{align}
In Secs.~\ref{sec:twist}--\ref{sec:model},
we will examine the properties of the $N$-th order Kohn--Drude weights in Eq.~\eqref{D_N}. 
We give concrete expressions of $\sigma_{(1)}^{L,\theta}(\omega)$, $\sigma_{(2)}^{L,\theta}(\omega)$, $\tilde{\mathcal{D}}_{(1)}^{L,\theta} $, and $\tilde{\mathcal{D}}_{(2)}^{L,\theta} $ below.

\subsection{Linear response}
According to Kubo's theory, the linear optical conductivity $\sigma_{(1)}^{L,\theta}(\omega)$ can be expressed as~\cite{Resta_2018,WatanabeLiuOshikawa}
\begin{align}
&\sigma_{(1)}^{L,\theta}(\omega)=\frac{i}{\omega+i\eta}\left\{\phi^{L,\theta}_{(1),0}(\omega)+\phi^{L,\theta}_{(1),1}(\omega)\right\},\label{eq:linear_optcond}
\end{align}
where $\phi^{L,\theta}_{(1),0}(\omega)=\langle\hat{k}^{L,\theta}\rangle$ and 
\begin{align}
    \phi^{L,\theta}_{(1),1}(\omega)&= L \sum_{n \neq 0} |\bra{0^{L,\theta}}\hat{j}^{L,\theta}\ket{n^{L,\theta}}|^2 \notag \\
    & \quad \times\left(\frac{1}{\omega - \Delta_n^{L,\theta} + i \eta} - \frac{1}{\omega + \Delta_n^{L,\theta} + i \eta} \right).
\end{align}
Here and hereafter, $\Delta_n^{L,\theta}\equiv E_n^{L,\theta}-E_0^{L,\theta}$ represents the excitation energy of the $n$-th eigenstate and  $\eta>0$ is an infinitesimal parameter.  The real part of $\sigma_{(1)}^{L,\theta}(\omega)$ can be decomposed into the singular and the regular parts:
\begin{align}
\mathrm{Re}[\sigma_{(1)}^{L,\theta}(\omega)]&=\pi\tilde{\mathcal{D}}_{(1)}^{L,\theta} \delta(\omega)+\mathrm{Re}[\sigma_{(1)\mathrm{reg}}^{L,\theta}(\omega)].
\end{align}
The coefficient of the first term
\begin{align}
\tilde{\mathcal{D}}_{(1)}^{L,\theta}& \equiv \langle\hat{k}^{L,\theta}\rangle - 2L \sum_{n > 0} \frac{|\langle n^{L,\theta}|\hat{j}^{L,\theta}|0^{L,\theta}\rangle|^2}{\Delta_{n}^{L,\theta}} \label{eq:linear_KohnDrude} 
\end{align}
gives the linear Kohn--Drude weight~\cite{Kohn1964}, for which Eq.~\eqref{D_N} with $N=1$ holds. The regular part
\begin{align}
\mathrm{Re}[\sigma_{(1)\mathrm{reg}}^{L,\theta}(\omega)]=\pi L \sum_{n>0}\frac{|\langle n^{L,\theta}|\hat{j}^{L,\theta}|0^{L,\theta}\rangle|^2}{\Delta_{n}^{L,\theta}}\delta(|\omega|-\Delta_{n}^{L,\theta})\label{eq:linear_opt_reg}
\end{align}
is a nonnegative and even function of $\omega$. The frequency-sum rule 
\begin{align}
\int_{-\infty}^{\infty}d\omega \sigma_{(1)}^{L,\theta}(\omega)=\int_{-\infty}^{\infty}d\omega \mathrm{Re}[\sigma_{(1)}^{L,\theta}(\omega)]=\pi\langle\hat{k}^{L,\theta}\rangle
\label{fsum}
\end{align}
and the positivity of the regular part implies that
\begin{align}
\label{Dkineq}
\tilde{\mathcal{D}}_{(1)}^{L,\theta}\leq \langle\hat{k}^{L,\theta}\rangle.
\end{align}

\subsection{Second order response}
Let us move to the second-order response. The second-order optical conductivity $\sigma_{(2)}^{L,\theta}(\omega_1,\omega_2)$ is given by
\begin{align}
\sigma_{(2)}^{L,\theta}(\omega_1,\omega_2)&=\frac{i}{\omega_1+i\eta}\frac{i}{\omega_2+i\eta}\sum_{j=0}^2\phi_{(2),j}^{L,\theta}(\omega_1, \omega_2)\label{eq:second_optcond},
\end{align}
where $\phi_{(2),0}^{L,\theta}(\omega_1, \omega_2)
=\langle\hat{t}^{L,\theta}\rangle$, 
\begin{widetext}
\begin{align}
&\phi_{(2),1}^{L,\theta}(\omega_1,\omega_2)
=L\sum_{n>0} \langle 0^{L,\theta}|\hat{j}^{L,\theta}|n^{L,\theta}\rangle\langle n^{L,\theta}|\hat{k}^{L,\theta}|0^{L,\theta}\rangle\left(
\frac{1}{\omega_1+\omega_2-\Delta_{n}^{L,\theta}+2i\eta}
-\frac{1}{\omega_1+\Delta_{n}^{L,\theta}+i\eta}
-\frac{1}{\omega_2+\Delta_{n}^{L,\theta}+i\eta}\right)\notag\\
&\quad\quad\quad-L\sum_{n>0}\langle 0^{L,\theta}|\hat{k}^{L,\theta}| n^{L,\theta}\rangle\langle n^{L,\theta}|\hat{j}^{L,\theta}|0^{L,\theta}\rangle\left(
\frac{1}{\omega_1+\omega_2+\Delta_{n}^{L,\theta}+2i\eta}
-\frac{1}{\omega_1-\Delta_{n}^{L,\theta}+i\eta}
-\frac{1}{\omega_2-\Delta_{n}^{L,\theta}+i\eta}\right),
\end{align}
and
\begin{align}
\phi_{(2),2}^{L,\theta}(\omega_1,\omega_2)
&=L^2\sum_{m,l > 0}\langle 0^{L,\theta}|\hat{j}^{L,\theta}|m^{L,\theta}\rangle\langle m^{L,\theta}|\hat{j}^{L,\theta}|l^{L,\theta}\rangle\langle l^{L,\theta}|\hat{j}^{L,\theta}|0^{L,\theta}\rangle \notag\\
&\qquad
\times
\Big[
\frac{1}{(\omega_1+\omega_2-\Delta_{m}^{L,\theta}+2i\eta)(\omega_1-\Delta_{l}^{L,\theta}+i\eta)}
+\frac{1}{(\omega_1+\omega_2-\Delta_{m}^{L,\theta}+2i\eta)(\omega_2-\Delta_{l}^{L,\theta}+i\eta)}\notag\\
&\qquad\qquad
-\frac{1}{(\omega_1+\Delta_{m}^{L,\theta}+i\eta)(\omega_2-\Delta_{l}^{L,\theta}+i\eta)}
-\frac{1}{(\omega_2+\Delta_{m}^{L,\theta}+i\eta)(\omega_1-\Delta_{l}^{L,\theta}+i\eta)} \notag\\
&\qquad\qquad
+\frac{1}{(\omega_1+\Delta_{m}^{L,\theta}+i\eta)(\omega_1+\omega_2+\Delta_{l}^{L,\theta}+2i\eta)}+\frac{1}{(\omega_2+\Delta_{m}^{L,\theta}+i\eta)(\omega_1+\omega_2+\Delta_{l}^{L,\theta}+2i\eta)}\Big]\notag\\
&\quad-L^2\sum_{n>0}\langle 0^{L,\theta}|\hat{j}^{L,\theta}|0^{L,\theta}\rangle\langle 0^{L,\theta}|\hat{j}^{L,\theta}|n^{L,\theta}\rangle\langle n^{L,\theta}|\hat{j}^{L,\theta}|0^{L,\theta}\rangle \notag\\
&\qquad
\times
\Big[\frac{1}{(\omega_1+\omega_2-\Delta_{n}^{L,\theta}+2i\eta)(\omega_1-\Delta_{n}^{L,\theta}+i\eta)}
+\frac{1}{(\omega_1+\omega_2-\Delta_{n}^{L,\theta}+2i\eta)(\omega_2-\Delta_{n}^{L,\theta}+i\eta)}\notag \\
&\qquad\qquad
-\frac{1}{(\omega_1+\Delta_{n}^{L,\theta}+i\eta)(\omega_2-\Delta_{n}^{L,\theta}+i\eta)}
-\frac{1}{(\omega_2+\Delta_{n}^{L,\theta}+i\eta)(\omega_1-\Delta_{n}^{L,\theta}+i\eta)}\notag\\
&\qquad\qquad
+\frac{1}{(\omega_1+\Delta_{n}^{L,\theta}+i\eta)(\omega_1+\omega_2+\Delta_{n}^{L,\theta}+2i\eta)}+\frac{1}{(\omega_2+\Delta_{n}^{L,\theta}+i\eta)(\omega_1+\omega_2+\Delta_{n}^{L,\theta}+2i\eta)}\Big].\label{eq:second_opt_phi2}
\end{align}
\end{widetext}
Extracting the coefficient of $\pi^2\delta(\omega_1)\delta(\omega_2)$ term, we obtain the second-order Kohn--Drude weight,
\begin{align}
&\tilde{\mathcal{D}}_{(2)}^{L,\theta} \equiv \langle\hat{t}^{L,\theta}\rangle\notag\\
& -6L\sum_{n>0} \frac{\mathrm{Re}[\langle 0^{L,\theta}|\hat{j}^{L,\theta}|n^{L,\theta}\rangle\langle n^{L,\theta}|\hat{k}^{L,\theta}|0^{L,\theta}\rangle]}{\Delta_{n}^{L,\theta}}\notag\\
&+6L^2\sum_{m,l>0} \frac{\langle 0^{L,\theta}|\hat{j}^{L,\theta}|m^{L,\theta}\rangle\langle m^{L,\theta}|\hat{j}^{L,\theta}|l^{L,\theta}\rangle\langle l^{L,\theta}|\hat{j}^{L,\theta}|0^{L,\theta}\rangle}{\Delta_{m}^{L,\theta}\Delta_{l}^{L,\theta}}\notag\\
&-6L^2\sum_{n>0}\frac{\bra{0^{L,\theta}}\hat{j}^{L,\theta}\ket{0^{L,\theta}}\langle 0^{L,\theta}|\hat{j}^{L,\theta}|n^{L,\theta}\rangle\langle n^{L,\theta}|\hat{j}^{L,\theta}|0^{L,\theta}\rangle}{\Delta_{n}^{L,\theta}\Delta_{n}^{L,\theta}}.\label{D_2}
\end{align}
This quantity can be written as Eq.~\eqref{D_N} with $N=2$~\cite{PhysRevB.102.165137}.

The above general formulas can be simplified for free fermions. We summarize these expressions in Appendix~\ref{TBexpress}. They require significantly less calculation cost and we will use them in our numerical demonstration.

\section{Twisted boundary condition and the ground state energy}
\label{sec:twist}
In this section we summarize the relation between the dependence of the ground state energy on the twisted boundary condition and the adiabatic transport of the system.

\subsection{General consideration}
\label{general}
We are interested in the $\theta$-dependence of the ground state energy $E_0^{L,\theta}$.
Since $e^{i\theta}$ may be interpreted as the phase of twisted boundary condition, $E_0^{L,\theta}$ must have the period $2\pi$ as a function of $\theta$. 
The $\theta$-dependence of the ground state energy  determines the ballistic transport property of the system.  
For example, the spontaneous current density is given by
\begin{align}
j_0^{L,\theta}\equiv \langle\hat{j}^{L,\theta}\rangle=\frac{dE_0^{L,\theta}}{d \theta}.
\label{jGS}
\end{align}
Higher-derivatives of $E_0^{L,\theta}$ are related to Drude weights as we have seen in Sec.~\ref{sec:review}.

The ground state energy $E_0^{L,\theta}$ can, in general, be expanded into a power series of $L$:
\begin{align}
E_0^{L,\theta}&=c_{+1}(\theta) L+c_0(\theta)+c_{-1}(\theta) L^{-1}+o(L^{-1}),
\label{GSEL}
\end{align}
where $o(L^{-n})$ represents corrections that decay faster than $L^{-n}$ in the large-$L$ limit, which includes possible terms with non-integer power $\alpha>1$ ($\alpha\notin\mathbb{Z}$) and logarithmic corrections. By definition, coefficients $c_p(\theta)$ ($p=1,0,-1,\cdots$) do not depend on $L$. Note that $c_{+1}(\theta)$ corresponds to the energy density $\varepsilon_0$ in the thermodynamic limit and cannot depend on $\theta$. 

In order to investigate the $\theta$-dependence of $c_0(\theta)$ and $c_{-1}(\theta)$, let us derive a bound for the spontaneous current density $j_0^{L,\theta}$ following the proof of the Bloch theorem~\cite{PhysRev.75.502,WatanabeBloch}. To this end, we introduce the large gauge transformation operator
\begin{equation}
\hat{U}\equiv e^{i(2\pi m/L)\sum_{x}x\hat{n}_x},
\end{equation}
where $\hat{n}_x$ is the number density operator at the site $x$. The operator $\hat{U}$ changes the flux $\theta$ by $2\pi m$, hence the gauge field $A=\theta/L$ by $2\pi m/L$.  Therefore,
\begin{align}
&\hat{U}^\dagger \hat{H}^{L,\theta}\hat{U}-\hat{H}^{L,\theta}\notag\\
&=\frac{2\pi m}{L}\frac{d\hat{H}^{L,\theta}}{d A}+\frac{1}{2}\left(\frac{2\pi m}{L}\right)^2\frac{d^2 \hat{H}^{L,\theta}}{d A^2}+O(L^{-2})\notag\\
&=2\pi m\hat{j}^{L,\theta}+2\pi^2 m^2L^{-1}\hat{k}^{L,\theta}+O(L^{-2}).
\end{align}
Here, $O(L^{-n})$ is a quantity that decays either equally fast with or faster than $L^{-n}$.
Since the variational principle implies $\langle\hat{U}^\dagger \hat{H}^{L,\theta} \hat{U}-\hat{H}^{L,\theta}\rangle\geq0$, we find
\begin{align}
mj_0^{L,\theta}+\pi m^2L^{-1}\langle \hat{k}^{L,\theta}\rangle+O(L^{-2})\geq0,
\end{align}
and hence
\begin{align}
m[\partial_\theta c_{0}(\theta)L+\partial_\theta c_{-1}(\theta) ]+\pi m^2\langle \hat{k}^{L,\theta} \rangle+o(1)\geq0
\label{mBloch}
\end{align}
for \emph{any} $m\in\mathbb{Z}$. It follows that $\langle \hat{k}^{L,\theta}\rangle \geq 0$ and $c_{0}(\theta)$ is 
$\theta$-independent. In fact, a non-zero $c_0(\theta)$ indicates the energy due to a defect, and thus $c_0(\theta)$ generally vanishes in translation-invariant systems. The statement $c_0(\theta)=0$ also follows from the conformal mapping from the infinite plane to a cylinder~\cite{BCN1986,Affleck-FSSCFT}, when the translation-invariant system is described by a CFT.

On the other hand, $c_{-1}(\theta)$ is related to the central charge and conformal dimensions of the CFT, and thus is expected to be universal (see the later sections for its precise meaning).
Nonvanishing $\theta$-dependence starts at $c_{-1}(\theta)$.
To quantify its $\theta$-dependence, let us look at the $n$-th derivative:
\begin{align}
d_{n}(\theta)\equiv\frac{d^{n}c_{-1}(\theta)}{d\theta^{n}}.
\label{dN}
\end{align}
These derivatives give the leading term of the spontaneous current density $j_0^{L,\theta}=d_1(\theta)L^{-1}+o(L^{-1})$ and the Kohn--Drude weights $\tilde{\mathcal{D}}_{(N)}^{L,\theta}=d_{N+1}(\theta)L^{N-1} + o(L^{N-1})$.
The inequality \eqref{mBloch} becomes
\begin{align}
md_{1}(\theta)+\pi m^2\langle \hat{k}^{L,\theta} \rangle+o(1)\geq0,
\end{align}
which implies
\begin{equation}
\left|d_{1}(\theta)\right|\leq\pi\lim_{L\to\infty}\langle\hat{k}^{L,\theta}\rangle=\lim_{L\to\infty}\int_{-\infty}^{\infty}d\omega \sigma_{(1)}^{L,\theta}(\omega).\label{newbound}
\end{equation}
The last equality is the frequency sum rule in Eq.~\eqref{fsum}. 
Furthermore, Eq.~\eqref{Dkineq} implies
\begin{equation}
d_2(\theta)\leq \lim_{L\to\infty}\langle\hat{k}^{L,\theta}\rangle
=\frac{1}{\pi}\lim_{L\to\infty}\int_{-\infty}^{\infty}d\omega \sigma_{(1)}^{L,\theta}(\omega).
\label{drudebound}
\end{equation}

\begin{figure}[t]
\begin{center}
\includegraphics[width=\columnwidth]{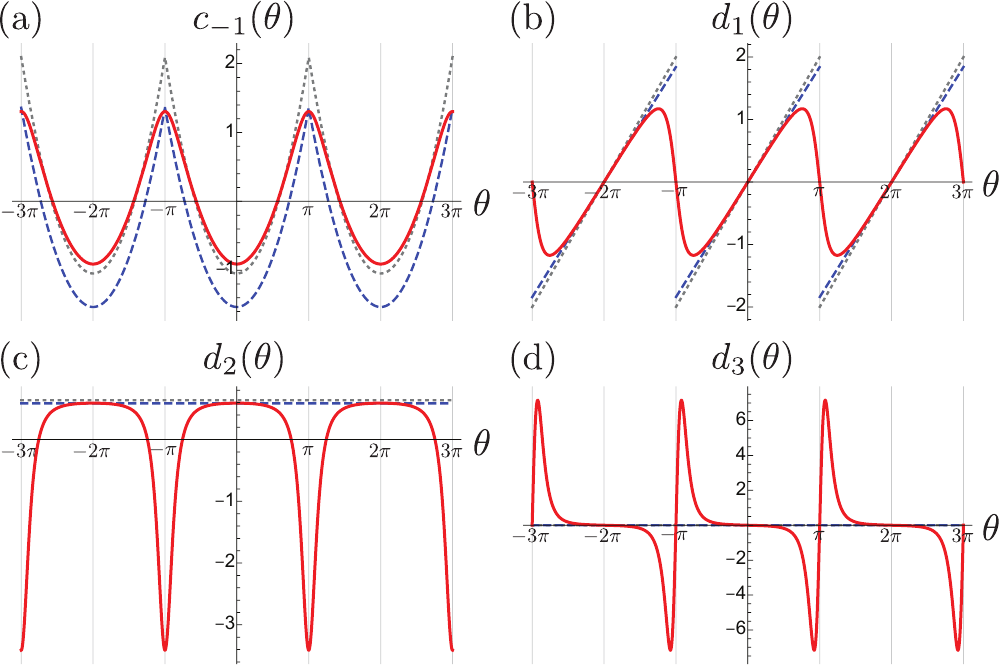}
\caption{\label{fig1}
(a) $c_{-1}(\theta)$ for the tight binding model ($t_0=1$) with single impurity potential 
$w=1$ (red solid line) and  $w=0$ (gray dotted line) at the half-filling. For comparison, we also plot $c_{-1}(\theta)$ for the XXZ model with $J=2$ and $\Delta=0.8$ (blue dashed line).
(b--d) The same as (a) but for $d_n(\theta)$. $n=1$ for (b), $n=2$ for (c), and $n=3$ for (d). 
}
\end{center}
\end{figure}

When $d_{N+1}(\theta)\neq0$ for an $N\geq2$, two important consequences follow immediately: (i) The $N$-th order Kohn--Drude weight $\tilde{\mathcal{D}}_{(N)}^{L,\theta}=d_{N+1}(\theta)L^{N-1} + o(L^{N-1})$ ($N\geq2$) diverges in the large-$L$ limit with the power $L^{N-1}$, and (ii) The \emph{linear} Kohn--Drude weight $\tilde{\mathcal{D}}_{(1)}^{L,\theta}=d_{2}(\theta) + o(1)$ depends nontrivially on $\theta$ even in the large-$L$ limit. This means 
that even the linear Kohn--Drude weight shows pathological behavior as a bulk quantity when $d_{n}(\theta)\neq0$ for an $n\geq3$. 
Since a bulk quantity in a sufficiently large system must be insensitive to the boundary condition, this implies that the linear Kohn--Drude weight defined in Eq.~\eqref{D_N} itself becomes ill-defined as a bulk quantity when $d_{n}(\theta)\neq0$ for an $n\geq3$. 
We will discuss the origin and the resolution of these behaviors in Sec.~\ref{sec:Dbulk}.
On the other hand, this suggests that $d_{n}(\theta)$ vanishes for all $n\geq3$ whenever the linear Kohn--Drude weight is supposed to be a well-defined bulk quantity independent of $\theta$. 
This is expected to be the case when the system has both U(1) symmetry and the lattice translation symmetry, regardless of the presence or the absence of many-body interactions. Even in this case the nonlinear Kohn--Drude weight $\tilde{\mathcal{D}}_{(N)}^{L,\theta}$ may still diverge but with a power smaller than $L^{N-1}$. This is the case of the XXZ model as we discuss later.

\subsection{Single-band tight-binding model}
\label{secH0}
As an example, let us consider a single-band tight-binding model of spinless electrons in one dimension. 
\begin{align}
\hat{H}_0^{L,\theta}&\equiv-t_0\sum_{x=-L/2+1}^{L/2}(\hat{c}_{x+1}^\dagger e^{-i\theta/L}\hat{c}_x+\mathrm{h.c.}).
\label{H0}
\end{align}
We set the lattice constant one and consider the PBC. We assume that the system size is $L=4\ell'+2$ and the number of electrons is $N_{\mathrm{el}}=2\ell+1$ ($\ell, \ell'\in\mathbb{N}$). 
The Hamiltonian can be diagonalized by the Fourier transformation $\hat{c}_{k_n}^\dagger\equiv L^{-1/2}\sum_{x}\hat{c}_{x}^\dagger e^{ik_n x}$. We get $\hat{H}_0^{L,\theta}=\sum_{n=1}^{L}\epsilon_{k_n}(\theta)\hat{c}_{k_n}^\dagger\hat{c}_{k_n}$ with $k_n\equiv 2\pi n/L$ and $\epsilon_{k_n}(\theta)=-2t_0\cos(k_n+\theta/L)$.  For the range $|\theta|<\pi$, the ground-state energy is given by
\begin{align}
E_0^{L,\theta}&=\sum_{n=-\ell}^{\ell}\epsilon_{k_n}(\theta)=-\frac{v_F}{\sin(\pi/L)}\cos(\theta/L).
\end{align}
Here the Fermi velocity $v_F$ is defined by
\begin{equation}
v_F\equiv 2t_0\sin k_F,\quad k_F\equiv\pi N_{\mathrm{el}}/L.\label{vF}
\end{equation}
Therefore,
\begin{align}
c_{+1}(\theta)&=-\frac{v_F}{\pi}\label{cp10},\\
c_0(\theta)&=0,\label{c010}\\
c_{-1}(\theta)&=\frac{v_F}{2\pi}[\arccos(\cos\theta)]^2-\frac{\pi v_F}{6}.\label{cm10}
\end{align}
This expression of $c_{-1}(\theta)$ respects the period $2\pi$ of $E_0^{L,\theta}$ and is valid for any $\theta\in\mathbb{R}$. 
Our convention of arccosine is the standard one, satisfying $0\leq\arccos(x)\leq\pi$, $\arccos(-1)=\pi$, and $\arccos(+1)=0$.  Thus $\arccos(\cos\theta)=|\theta|$ for $|\theta|<\pi$. 
We plot $c_{-1}(\theta)$ and its derivative $d_{n}(\theta)$ ($n=1, 2, 3$) of this model in Fig.~\ref{fig1}.

The $N$-th order Kohn--Drude weight in this model remains finite in the thermodynamic limit, which is given by $\lim_{L\to\infty}\tilde{\mathcal{D}}_{(N)}^{L,\theta}=(-1)^{(N-1)/2}v_F/\pi$ when $N$ is odd and $\lim_{L\to\infty}\tilde{\mathcal{D}}_{(N)}^{L,\theta}=0$ when $N$ is even.
More generally, all nonlinear Kohn--Drude weights remain finite if the ground state energy takes the form
\begin{align}
E_0^{L,\theta}&=L\varepsilon_0^L(\theta/L)
\end{align}
with a function $\varepsilon_0^L(A)$ whose large-$L$ limit $\varepsilon_0(A)$ is a smooth function of $A$. In the above tight-binding model, $\varepsilon_0(A)=-(v_F/\pi)\cos A$. In such a case, $c_{p}(\theta)$ ($p=1,0,-1,\cdots$) is a polynomial of $\theta$ with the maximal power $\theta^{1-p}$ and $\lim_{L\to\infty}\tilde{\mathcal{D}}_{(N)}^{L,\theta}=d^{N+1}\varepsilon_{0}(A)/dA^{N+1}|_{A=0}$.

\subsection{Tomonaga-Luttinger liquids}
\label{XXZ}
For Tomonaga-Luttinger liquids (TLLs) with the Luttinger parameter $K$ and the velocity parameter $v$, the finite-size scaling of the ground state energy is known to be~\cite{Giamarchi2004B}
\begin{align}
c_{+1}(\theta)&=\varepsilon_0,\\
c_0(\theta)&=0,\\
c_{-1}(\theta)&=                              
\frac{K v}{2\pi} [\arccos(\cos\theta)]^2
- \frac{\pi v}{6}.
\label{cmTLL}
\end{align}
For example, in the case of the spin-1/2 XXZ Heisenberg spin chain
\begin{equation}
\hat{H}^{L,\theta}_\mathrm{XXZ}=J\sum_{x=1}^L\left(\frac{1}{2}\hat{s}_{x+1}^+e^{-i\theta/L}\hat{s}_x^-+\mathrm{h.c.}+\Delta\hat{s}_{x+1}^z\hat{s}_x^z\right)
\end{equation} with $-1<\Delta\leq1$, 
the parameters $\varepsilon_0$, $K$, and $v$ are given by~\cite{Hamer_1987,Sutherland1990,doi:10.1142/S0217979212440092}
\begin{align}
\varepsilon_0&=\frac{J}{4}\cos\gamma-\frac{J}{2}\sin\gamma\int_{-\infty}^\infty dx\frac{\sinh[(\pi-\gamma) x]}{\sinh(\pi x)\cosh(\gamma x)},\\
v &= J\frac{\pi\sin{\gamma}}{2\gamma},\quad K =\frac{\pi}{2(\pi-\gamma)},
\end{align}
where $\gamma\equiv\arccos\Delta$. 
This model with $J=2t_0$ and $\Delta=0$ can be exactly mapped to the above tight-binding model at the half-filling $N_{\mathrm{el}}=L/2$.
We plot $c_{-1}(\theta)$ and its derivative $d_{n}(\theta)$ ($n=1, 2, 3$) in Fig.~\ref{fig1}. 

Although TLLs satisfy $d_{n}=0$ for $n\geq3$, 
nonlinear Kohn--Drude weights diverge in the thermodynamic limit
in the XXZ chain as studied in Refs.~\cite{PhysRevB.102.165137, Tanikawa2021, Fava}. This is due to the irrelevant perturbation to the CFT and not captured in the unperturbed TLLs~\cite{Tanikawa2021}. We emphasize that this fact does not contradict our statement because the divergence of the $N$-th order Kohn--Drude weight in the XXZ chain is with a power smaller than $L^{N-1}$.

We observe that $c_{-1}(\theta)$ in Eq.~\eqref{cmTLL} is written completely in terms of the parameters of the low-energy effective theory and is universal in that sense.
In particular, the constant term $-\pi v/6$ follows from the central charge $1$ of the TLL.
The Luttinger parameter $K$ only affects the coefficient of $\theta^2$ but never generates higher power terms of $\theta$ in $c_{-1}(\theta)$. Reflecting the level crossing of many-body energy levels at $\theta=(2m-1)\pi$ ($m\in\mathbb{Z}$), the slope of $c_{-1}(\theta)$ is discontinuous at these points.  The level crossing is protected by the lattice translation symmetry; that is, the two many-body energy levels crossing at these points have distinct momenta ($0$ and $2k_F$) and they cannot repel each other. This observation motivates us to investigate translation-breaking perturbations.

\section{Tight-binding model with a single defect}
\label{sec:model}
Let us introduce a defect $\hat{V}$ to the tight-binding model $\hat{H}_0^{L,\theta}$ in Sec.~\ref{secH0}. The defect  induces a level repulsion at $\theta=(2m-1)\pi$ and $E_0^{L,\theta}$ becomes a smooth periodic function of $\theta$. As a consequence, the $\theta$-dependence of $c_{-1}(\theta)$ is fundamentally modified, as we shall see in Sec.~\ref{GSE}.

\subsection{Single defect scattering}
We consider a defect $\hat{V}$ localized around the site $x=0$.
Examples include a single impurity potential
\begin{equation}
\hat{V}=w\hat{c}_0^\dagger\hat{c}_0
\label{impuritypotential}
\end{equation}
and a single bond disorder
\begin{equation}
\hat{V}=-(ve^{i\delta}-t_0)\hat{c}_{1}^\dagger e^{-i\theta/L}\hat{c}_{0}+\mathrm{h.c.},
\label{bonddisorder}
\end{equation}
but our discussion below is not restricted to these cases. The only assumptions are that $\hat{V}$ is written in terms of operators near the origin and is a bilinear of $\hat{c}_x^\dagger$ and $\hat{c}_{x'}$.  

To solve the eigenvalue equation $(\hat{H}_0^{L,\theta}+\hat{V})|k\rangle=\epsilon_k|k\rangle$, 
we postulate the following form of the wavefunction:
\begin{align}
\psi_k(x)\equiv\tilde{\psi}_k^+e^{i kx}+\tilde{\psi}_k^-e^{-i (k+2\theta/L)x}\label{pw1}
\end{align}
for $1-L/2\leq x\ll-1$ and
\begin{align}
\psi_k(x)\equiv\psi_k^+e^{i kx}+\psi_k^-e^{-i (k+2\theta/L)x}\label{pw2}
\end{align}
for $1\ll x \leq L/2$. We may assume $0<k+\theta/L<\pi$. The energy eigenvalue is still given by $\epsilon_k=-2t_0\cos(k+\theta/L)$ as in the defect-free case, but the quantization condition imposed on $k$ is modified, as we will see below.

The defect $\hat{V}$ can be characterized by the scattering matrix
\begin{align}
S_{q}\equiv
\begin{pmatrix}
T_{q}^+&R_{q}^-\\
R_{q}^+&T_{q}^-
\end{pmatrix},
\end{align}
where $T_q^\pm$ and $R_q^\pm$ are the transmission and the reflection coefficients. It maps the incoming components ($\tilde{\psi}_k^+$ and $\psi_k^-$) to the outgoing components ($\psi_k^+$ and $\tilde{\psi}_k^-$):
\begin{align}
\begin{pmatrix}
\psi_k^+\\
\tilde{\psi}_k^-
\end{pmatrix}=S_{k+\theta/L}
\begin{pmatrix}
\tilde{\psi}_k^+\\
\psi_k^-
\end{pmatrix}.
\label{condition1}
\end{align}
For example, for the single impurity potential~\eqref{impuritypotential},
\begin{align}
T_q^\pm=\frac{2t_0\sin q}{2t_0\sin q+iw},\quad
R_q^\pm=-\frac{iw}{2t_0\sin q+iw}.
\end{align}
For the single bond disorder~\eqref{bonddisorder},
\begin{align}
&T_q^\pm=e^{\pm i\delta}\frac{2t_0v\sin q}{(t_0^2+v^2)\sin q+i(t_0^2-v^2)\cos q},\\
&R_q^\pm=-e^{\pm iq}\frac{i(t_0^2-v^2)}{(t_0^2+v^2)\sin q+i(t_0^2-v^2)\cos q}.
\end{align}

The conservation of the probability current implies that the $S$-matrix is unitary for any $q$.
Furthermore, it is also constrained by the symmetries of the system. For example, the time-reversal invariance implies
\begin{align}
\begin{pmatrix}
\psi_k^+\\
\tilde{\psi}_k^-
\end{pmatrix}
=
\sigma_1
\begin{pmatrix}
\tilde{\psi}_k^+\\
\psi_k^-
\end{pmatrix}^*,
\end{align}
where $\sigma_i$ ($i=1,2,3$) is the Pauli matrix.
Together with the unitarity, we find $\sigma_1 S_q^* \sigma_1 = S_q^\dagger$, which is reduced to $T_q^+ = T_q^-$~\cite{doi:10.1063/1.531762}.
Likewise, the spatial inversion about $x=x_0$ requires
\begin{align}
\begin{pmatrix}
\tilde{\psi}_k^+\\
\psi_k^-
\end{pmatrix} &= \sigma_1 e^{2ix_0q\sigma_3} \begin{pmatrix}
\tilde{\psi}_k^+\\
\psi_k^-
\end{pmatrix},\\
\begin{pmatrix}
\psi_k^+\\
\tilde{\psi}_k^-
\end{pmatrix} &= \sigma_1 e^{2ix_0q\sigma_3}\begin{pmatrix}
\psi_k^+\\
\tilde{\psi}_k^-
\end{pmatrix},
\end{align}
implying $S_q = e^{-2iqx_0\sigma_3}\sigma_1 S_q \sigma_1 e^{2iqx_0\sigma_3}$. In terms of the matrix elements, this is equivalent to
$T_q^+ =  T_q^-$ and $R_q^- =e^{-4iqx_0}  R_q^+$. For example, the impurity potential~\eqref{impuritypotential} has the site inversion symmetry ($x_0=0$) and the bond disorder~\eqref{bonddisorder} has the bond inversion symmetry ($x_0=1/2$) when $\delta=0$.
Note that our discussions below do not assume any of these symmetries.

\subsection{Quantization condition}
Now we impose the PBC:
\begin{align}
\begin{pmatrix}
\tilde{\psi}_k^+\\
\tilde{\psi}_k^-
\end{pmatrix}=
\begin{pmatrix}
\psi_k^+e^{ikL}\\
\psi_k^-e^{-i(k+2\theta/L)L}
\end{pmatrix}.
\label{condition2}
\end{align}
We demand the existence of nonvanishing solutions to Eqs.~\eqref{condition1} and \eqref{condition2}, which leads to the quantization condition of $q\equiv k+\theta/L$. Such a condition can be most easily implemented (see Ref.~\cite{Oshikawa-IsingGap} for a related discussion on Majorana fermions) by parametrizing the scattering matrix as
\begin{align}
S_q=e^{i\varphi_q}
\begin{pmatrix}
T_qe^{i\delta_q}&-R_qe^{-i\eta_q}\\
+R_qe^{i\eta_q}&T_qe^{-i\delta_q}
\end{pmatrix},
\label{parametrization}
\end{align}
where $T_q\equiv|T_q^\pm|$ and $R_q\equiv|R_q^\pm|$ are the transmission and the refection amplitude. The phase $\varphi_q$ is related to the determinant of the scattering matrix as $\det S_q=T_q^+/(T_q^-)^*=e^{2i\varphi_q}$. In the presence of either the time-reversal symmetry or the inversion symmetry, $\delta_q=0$.  With these definitions, the quantization condition reads
\begin{align}
\cos(qL+\varphi_{q})=T_q\cos(\theta-\delta_{q}),\quad 0<q<\pi.
\label{quantization}
\end{align}

The parametrization~\eqref{parametrization} and the quantization condition  \eqref{quantization} are invariant under the simultaneous shift $\varphi_q\to \varphi_q+\pi$, $\delta_q\to \delta_q+\pi$, and $\eta_q\to\eta_q+\pi$. To fix the ambiguity, here we assume the absence of bound states below the band bottom $\epsilon=-2t_0$.
We choose a branch of $\varphi_{q}$ in such a way that $\lim_{q\to+0}\varphi_{q}=-\pi/2$~\cite{doi:10.1063/1.531762} and is continuous as a function of $q$. See Sec.~\ref{boundstate} for the case when bound states appear below $-2t_0$.

In the defect-free case (i.e., $T_q=1$ and $\varphi_q=\delta_q=0$), the solutions to Eq.~\eqref{quantization} can be written as
\begin{align}
q_n^\pm=k_n\pm\frac{|\theta|}{L}\quad \left(k_n\equiv \frac{2\pi n}{L}\right),
\label{w0}
\end{align}
where $n=0,1,\cdots,L/2-1$ for $q_n^+$ and $n=1,2,\cdots,L/2$ for $q_n^-$. Even for a general $\hat{V}\neq0$, Eq.~\eqref{quantization} can be expressed in a form similar to Eq.~\eqref{w0}:
\begin{align}
&q_n^{\pm}=k_n+\frac{\phi_\pm(q_n^{\pm})}{L},\label{wp}\\
&\phi_\pm(q)\equiv\pm\arccos\left(T_q\cos(\theta-\delta_q)\right)-\varphi_q.\label{phipm}
\end{align}
Since the phase shift $\phi_\pm(q_n^\pm)$ in Eq.~\eqref{wp} depends on $q_n^\pm$, it still needs to be solved self-consistently.
In practice, however, one can solve it iteratively. Namely, the first approximation is to replace $q_n^\pm$ on the right hand side by $k_n$, which gives $q_n^\pm$ with an error $O(L^{-2})$. For our purpose of determining $c_{-1}(\theta)$, one needs to repeat this step once again to determine $q_n^\pm$ to the $L^{-2}$ accuracy.  We find
\begin{align}
q_n^\pm=k_n+\frac{\phi_\pm(k_n)}{L}+\frac{1}{2L^2}\frac{d[\phi_\pm(k_n)]^2}{dk_n}+O(L^{-3}).
\label{solq}
\end{align}

\subsection{Ground state energy}
\label{GSE}
Using Eq.~\eqref{solq} and performing the Taylor expansion in a series of $L^{-n}$, we evaluate the ground state energy as
\begin{align}
E_0^{L, \theta}&=-2t_0\sum_{n=0}^{\ell}\cos(q_n^+)-2t_0\sum_{n=1}^{\ell}\cos(q_n^-)\notag\\
&=\tilde{c}_{+1}^L(\theta)L+\tilde{c}_0^{L,\theta}(\theta)+\tilde{c}_{-1}^L(\theta)L^{-1}+O(L^{-2}).
\end{align}
Here, the coefficients $\tilde{c}_{p}^L(\theta)$ are given by
\begin{align}
\tilde{c}_{+1}^L(\theta)&\equiv-\frac{2t_0}{L}\sum_{n=-\ell}^{\ell}\cos k_n=-\frac{v_F}{\pi}-\frac{\pi v_F}{6L^2}+O(L^{-4}),\label{cpt1}
\end{align}
\begin{align}
\tilde{c}_0^{L}(\theta)&\equiv\frac{2t_0}{L}\sum_{n=1}^{\ell}\sin k_n\left[\phi_+(k_n)+\phi_-(k_n)\right]\notag\\
&=-\frac{2t_0}{\pi}\int_0^{k_F}dk\sin k\,\varphi_k+O(L^{-2}),
\end{align}
and
\begin{align}
\tilde{c}_{-1}^L(\theta)
&\equiv\frac{t_0}{L}\sum_{n=0}^{\ell}\frac{d}{dk_n}\left[\sin(k_n)\phi_+(k_n)^2\right]\notag\\
&\quad+\frac{t_0}{L}\sum_{n=1}^{\ell}\frac{d}{dk_n}\left[\sin(k_n)\phi_-(k_n)^2\right]\notag\\
&=\frac{v_F}{4\pi}[\phi_+(k_F)^2+\phi_-(k_F)^2]+O(L^{-1}).\label{cmt1}
\end{align}
Combining these results with Eq.~\eqref{phipm}, we find
\begin{align}
c_{+1}(\theta)&=-\frac{v_F}{\pi},\label{cp}\\
c_{0}(\theta)&=-\frac{2t_0}{\pi}\int_0^{k_F}dk\sin k\,\varphi_k,\label{c0}\\
c_{-1}(\theta)&=\frac{v_F}{2\pi}\left[\arccos\left(T_F\cos(\theta-\delta_F)\right)\right]^2+\frac{v_F}{2\pi}\varphi_{F}^2-\frac{\pi v_F}{6}.\label{cm}
\end{align}
In these expressions, $v_F$ and $k_F$ are defined in Eq.~\eqref{vF}.
In the derivation, we used the Euler--Maclaurin formula $\sum_{n=1}^{m}f(k_n)=L/(2\pi)\int_{k_1}^{k_{m+1}}dkf(k)-(1/2)[f(k_{m+1})-f(k_1)]+O(L^{-1})$ for the error estimate. The subscript $F$ refers to the value at the Fermi point $k=k_F$. The effect of $\delta_F$ merely shifts the origin of $\theta$.

We note that the present result is consistent with the earlier general observation.
The ``defect energy'' $c_0(\theta)$ is nonvanishing only in the presence of the localized defect.
On the other hand, $c_{-1}(\theta)$ is written completely in terms of the parameters of the low-energy effective theory, including the transmission amplitude $T_F$ at the Fermi point. The constant term $-\pi v_F/6$ is again the consequence of the central charge $1$ of the corresponding CFT (TLL). In this sense $c_{-1}(\theta)$ is universal. For more detail, see Appendix~\ref{Sec:BCFT}.

\subsection{Bound states}
\label{boundstate}
Some of eigenstates of $\hat{H}_0^{L,\theta}+\hat{V}$ are exponentially localized to the defect. These bound states have energy eigenvalues lower than $-2t_0$ or higher than $+2t_0$. The number of plane-wave states in Eqs.~\eqref{pw1} and \eqref{pw2} is reduced by the number of the bound states. For example, the impurity potential in Eq.~\eqref{impuritypotential} has a bound state when $|w|L> 1-\cos\theta$. The energy eigenvalue can be approximated by
$\mathrm{sign}(w)\,\sqrt{(2t_0)^2+w^2}$ in a sufficiently large system~\footnote{More precisely, one needs to find $\lambda>0$ by solving $2t_0(\cosh L\lambda-\cos\theta)\sinh\lambda/\sinh L\lambda=|w|$. The energy eigenvalue is given by $\mathrm{sign}(w)2t_0\cosh\lambda$.}.

Let us assume that 
there are $N_{\mathrm{b}}$ bound states below $\epsilon=-2t_0$. We write their energy eigenvalues as $\epsilon_m^{\mathrm{b}}$ ($m=1,2\cdots,N_{\mathrm{b}}$). 
In this situation, we find it useful to define
\begin{equation}
\tilde{\varphi}_q\equiv\varphi_q-N_{\mathrm{b}}\pi.
\end{equation}
The Levinson theorem states $\lim_{q\to+0}\tilde{\varphi}_{q}=-\pi/2$~\cite{doi:10.1063/1.531762}. 
Our results in Eqs.~\eqref{c0} and \eqref{cm} are modified as
\begin{align}
c_{0}(\theta)&
=-\frac{2t_0}{\pi}\int_0^{k_F}dk\sin k\,(\tilde{\varphi}_k+N_{\mathrm{b}}\pi)+\sum_{m=1}^{N_{\mathrm{b}}}(\epsilon_m^{\mathrm{b}}+2t_0)\notag\\
&=-\frac{2t_0}{\pi}\int_0^{k_F}dk\sin k\,\tilde{\varphi}_k+\sum_{m=1}^{N_{\mathrm{b}}}(\epsilon_m^{\mathrm{b}}-\epsilon_F),\label{c0bp}\\
c_{-1}(\theta)&=\frac{v_F}{2\pi}\left[\arccos\left(T_F\cos(\theta-\delta_F)\right)\right]^2\notag\\
&\quad\quad+\frac{v_F}{2\pi}(\tilde{\varphi}_{F}+N_{\mathrm{b}}\pi)^2-\frac{\pi v_F}{6}.\label{cmbp}
\end{align}
The first term of $c_{-1}(\theta)$, which governs the $\theta$-dependence, is independent of the bound states away from the Fermi level. The second term weakly depends on them through the choice of the branch of $\varphi_q$. This is in contrast to the non-universal defect energy $c_0(\theta)$, which explicitly depends on the bound state energy $\epsilon_m^{\mathrm{b}}$.

\begin{figure*}[t]
    \begin{center}
    \includegraphics[width=17.5cm]{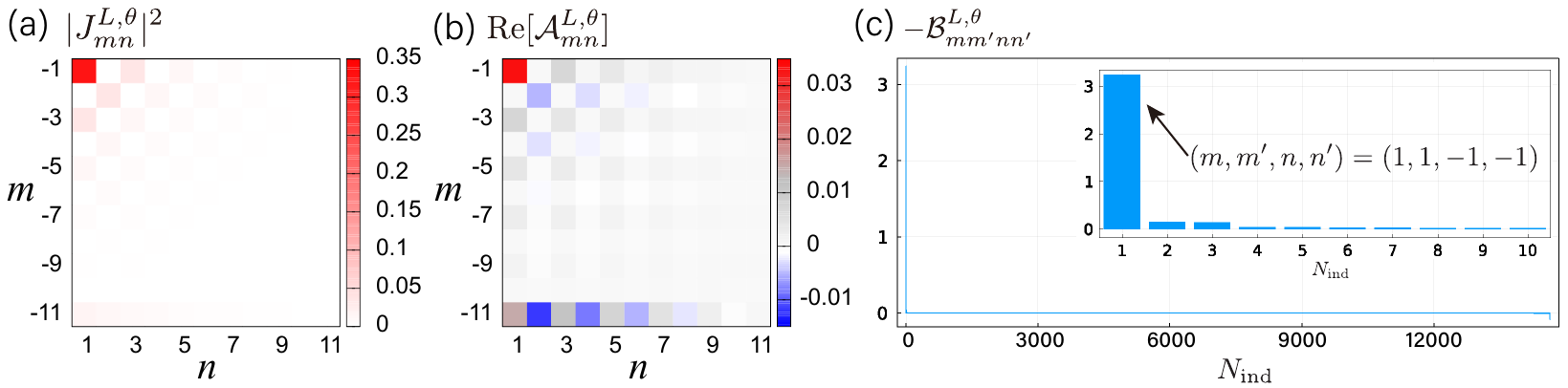}
    \caption{
    (a) $|J^{L,\theta}_{mn}|^2$ for $\theta=1.57$ and $L=22$. (b) $\mathrm{Re}[\mathcal{A}_{mn}]$ for $\theta=2.34$.
    (c) $-\mathcal{B}_{mm^\prime n n^\prime}$ for $\theta=2.34$ and $L=22$. We plot all the combination of $(m,m^\prime,n,n^\prime)$ for $m,m'>0$ and $n,n'<0$. The horizontal axis $N_\mathrm{ind}$ denotes the $N_\mathrm{ind}$-th largest value of $\mathcal{B}_{mm^\prime n n^\prime}$. The data from the largest to the 10-th largest one are shown in the inset. We set $t_0=w=1$.
    \label{fig:matele}
    }
    \end{center}
    \end{figure*}

\subsection{Derivatives of the ground state energy}
Given the fully general expression of $c_{-1}(\theta)$ in Eq.~\eqref{cm}, we can readily compute the derivatives of $c_{-1}(\theta)$ in Eq.~\eqref{dN}. For example,
\begin{align}
&d_1(\theta)=\frac{v_F\sin\theta'\arccos(T_F\cos\theta')}{\pi\sqrt{\sin^2\theta'+r_F^2}},
\label{d1}
\end{align}
\begin{align}
&d_2(\theta)=\frac{v_F}{\pi}\frac{\sin^2\theta'}{\sin^2\theta'+r_F^2}+d_1(\theta)\frac{r_F^2\cot\theta'}{\sin^2\theta'+r_F^2},
\label{d2}
\end{align}
and
\begin{align}
&d_3(\theta)=\frac{v_F}{2\pi}\frac{3r_F^2\sin2\theta'}{(\sin^2\theta'+r_F^2)^2}-d_1(\theta)\frac{r_F^2(2+\cos2\theta'+r_F^2)}{(\sin^2\theta'+r_F^2)^{2}},
\label{d3}
\end{align}
where $\theta'\equiv\theta-\delta_F$ and $r_F\equiv R_F/T_F$. 
Note that Eqs.~\eqref{cm} and \eqref{d1} were previously derived in Ref.~\cite{Gogolin1994} for the special case of an impurity potential in a continuum model. Here we rederived it in a more general setting on the lattice model without specifying the form of the impurity $\hat{V}$. Possible time-reversal symmetry breaking by $\hat{V}$ results in nonzero $\delta_F$ in our setting.
We plot these functions in Fig.~\ref{fig1} for the example of the single impurity potential at the half-filling, for which we have $r_F=|w|/v_F$ and $v_F=2t_0$.  We see that $d_{n}(\theta)$ generally does not vanish, implying the divergence of $\mathcal{D}_{(n-1)}^{L,\theta}$ in the thermodynamic limit, except for several special values of $\theta$. For example, the time-reversal symmetry implies $d_{2n-1}(\theta)=0$ for $\theta=0$ and $\pi$, but $d_1(\theta),d_3(\theta)\neq0$ except for these points.  Despite this divergence, $|d_1(\theta)|$, which is the leading term in the adiabatic current $j_0^{L,\theta}=d_1(\theta)L^{-1}+o(L^{-1})$, is a monotonically decreasing function of $r_F$ for a fixed $\theta$. In fact, for the present model, the frequency sum of the optical conductivity $\sigma_{(1)}^{L,\theta}(\omega)$ in Eq.~\eqref{newbound} is given by
\begin{align}
\pi \langle\hat{k}^{L,\theta} \rangle=-\frac{\pi}{L}\langle\hat{H}^{L,\theta}\rangle+O(L^{-1})=v_F+O(L^{-1}),
\end{align}
implying that $|d_1(\theta)|\leq v_F$ regardless of $\theta$ or $\hat{V}$ [see Eq.~\eqref{newbound}]. The bound is saturated at $\theta=\pm\pi$, in the absence of the defect. Therefore, at least in this class of models, the diverging Kohn--Drude weights do not imply any larger adiabatic current.

The $\theta$-dependence of $d_n(\theta)$ may be related to the singularity around $\theta'=\theta-\delta_F=\pi$ and $r_F=0$:
\begin{align}
d_{2}(\pi+\delta_F)&=d_{2}(\delta_F)-\frac{v_F}{r_F},\label{s2}\\
d_{4}(\pi+\delta_F)&=d_{4}(\delta_F)+v_F\left(\frac{3}{r_F^3}+\frac{1}{r_F}\right),\label{s4}
\end{align}
which diverges in the $r_F=0$ limit. This divergence originates from the degeneracy of the single-particle levels of $\hat{H}_0^{L,\theta}$ at $\theta=\pi$ as we show in Appendix~\ref{perturbation}. It is interesting to observe that the absolute value of the linear Kohn--Drude weight can become arbitrarily large without contradicting with the bound $d_2(\theta)\leq v_F/\pi$ that follows from Eq.~\eqref{drudebound}.

\subsection{Response theory and divergence}

We have seen the divergence of the Kohn--Drude weights using the Kohn formula. Here, we show that the divergence can also be explained by the response theory formulas [Eqs.~\eqref{eq:linear_KohnDrude} and \eqref{D_2}].

For this purpose, we use the expressions of the Kohn--Drude weights for tight-binding models given in Appendix~\ref{TBexpress}. The linear and second-order Kohn--Drude weights are given by
\begin{align}
    \tilde{\mathcal{D}}_{(1)}^{L,\theta} &= \frac{1}{L}\sum_{n < 0} K_{nn}^{L,\theta}  - \frac{2}{L}\sum_{\substack{n < 0,\\ m > 0}} \frac{|J_{mn}^{L,\theta}|^2}{\epsilon_{mn}^{L,\theta}}, \label{eq:linear_Kohn_Drude_rewrite} \\
    \tilde{\mathcal{D}}_{(2)}^{L,\theta} &=  \frac{1}{L}\sum_{n < 0} T^{L,\theta}_{nn} -\frac{6}{L}\sum_{\substack{n < 0, \\ m > 0}} \frac{\mathrm{Re}[\mathcal{A}_{mn}^{L,\theta}]}{\epsilon^{L,\theta}_{mn}}+\frac{6}{L}\sum_{\substack{n,n'<0,\\m,m'>0}} \frac{\mathcal{B}_{m m^\prime n n^\prime}^{L,\theta}}{\epsilon^{L,\theta}_{mn}\epsilon^{L,\theta}_{m'n'}} \label{eq:second_Kohn_Drude_rewrite},
\end{align}        
where we wrote the matrix elements of $\hat{j}^{L,\theta}$, $\hat{k}^{L,\theta}$, and $\hat{k}^{L,\theta}$ among single-particle states as $J_{mn}^{L,\theta}/L$, $K_{mn}^{L,\theta}/L$, and $T_{mn}^{L,\theta}/L$, respectively, and we defined $\mathcal{A}_{mn}^{L,\theta}\equiv J_{nm}^{L,\theta} K_{mn}^{L,\theta}$ and $\mathcal{B}_{m m^\prime n n^\prime}^{L,\theta} \equiv J^{L,\theta}_{nm} (J^{L,\theta}_{mm'} \delta_{nn'} - J^{L,\theta}_{n'n} \delta_{mm'}) J^{L,\theta}_{m'n'}$. As shown in Appendix~\ref{App:finiteness}, $J_{mn}^{L,\theta}$, $K_{mn}^{L,\theta}$, and $T_{mn}^{L,\theta}$ are $O(1)$.
In contrast, the energy difference of the two single particle states $\epsilon_{mn}^{L,\theta}\equiv \epsilon_m^{L,\theta}-\epsilon_n^{L,\theta}$ can be $O(1/L)$.
Using these properties, and counting the orders, we find that there is no divergent term at the linear order [Eq.~\eqref{eq:linear_Kohn_Drude_rewrite}]. At the second order, the third term in Eq.~\eqref{eq:second_Kohn_Drude_rewrite} is proportional to $L$ and diverges in the large-$L$ limit due to the two factors of $\epsilon^{L,\theta}_{mn}$ in the denominator. More generally, $\epsilon^{L,\theta}_{mn}$ appears $N$ times in the denominator at the $N$-th order and $\tilde{\mathcal{D}}_{(N)}^{L,\theta}$ is proportional to $L^{N-1}$. These observations are consistent with our results based on the Kohn formulas.

To see the behavior of $\tilde{\mathcal{D}}_{(1)}^{L,\theta}$ and $\tilde{\mathcal{D}}_{(2)}^{L,\theta}$ in more detail, we study the structure of the matrix elements $J_{mn}^{L,\theta}$, $\mathcal{A}_{mn}^{L,\theta}$, and $\mathcal{B}_{m m^\prime n n^\prime}^{L,\theta}$. As shown in Fig.~\ref{fig:matele}, the matrix elements have peak structures. $J_{mn}^{L,\theta}$, $\mathcal{A}_{mn}^{L,\theta}$ and $\mathcal{B}_{mm^\prime nn^\prime}^{L,\theta}$ take peaks at $(m,n)=(1,-1)$, $(m,n)=(1,-1)$, and $(m,m^\prime,n,n^\prime)=(1,1,-1,-1)$ respectively. Thus, we can approximate the Kohn--Drude weights as
\begin{align}
    \tilde{\mathcal{D}}_{(1)}^{L,\theta}&\simeq \langle \hat{k}^{L,\theta} \rangle - 2c|J^{L,\theta}_{1,-1}|^2, \label{eq:D1approx} \\
    \tilde{\mathcal{D}}_{(2)}^{L,\theta}&\simeq \langle \hat{t}^{L,\theta} \rangle - 6c \mathrm{Re}[\mathcal{A}^{L,\theta}_{1,-1}] + 6 L c^2 \mathcal{B}^{L,\theta}_{1,1,-1,-1}, \label{eq:D2approx}
\end{align}
where an order-one constant $c$ is defined as $\epsilon^{L,\theta}_{1,-1}=c/L$. This clearly shows $\tilde{\mathcal{D}}_{(2)}^{L,\theta} \propto L$ for large $L$. If the defect potential is absent ($w=0$), $J_{mn}^{L,\theta}$, $\mathcal{A}_{mn}^{L,\theta}$ and $\mathcal{B}_{mm^\prime nn^\prime}^{L,\theta}$ becomes zero for $m>0$ and $n<0$ and thus the Kohn--Drude weights take the $f$-sum values such as $\langle \hat{k}^{L,\theta} \rangle$ and $\langle \hat{t}^{L,\theta} \rangle$~\cite{PhysRevB.102.165137}.

\begin{figure*}[t]
    \begin{center}
    \includegraphics[width=17cm]{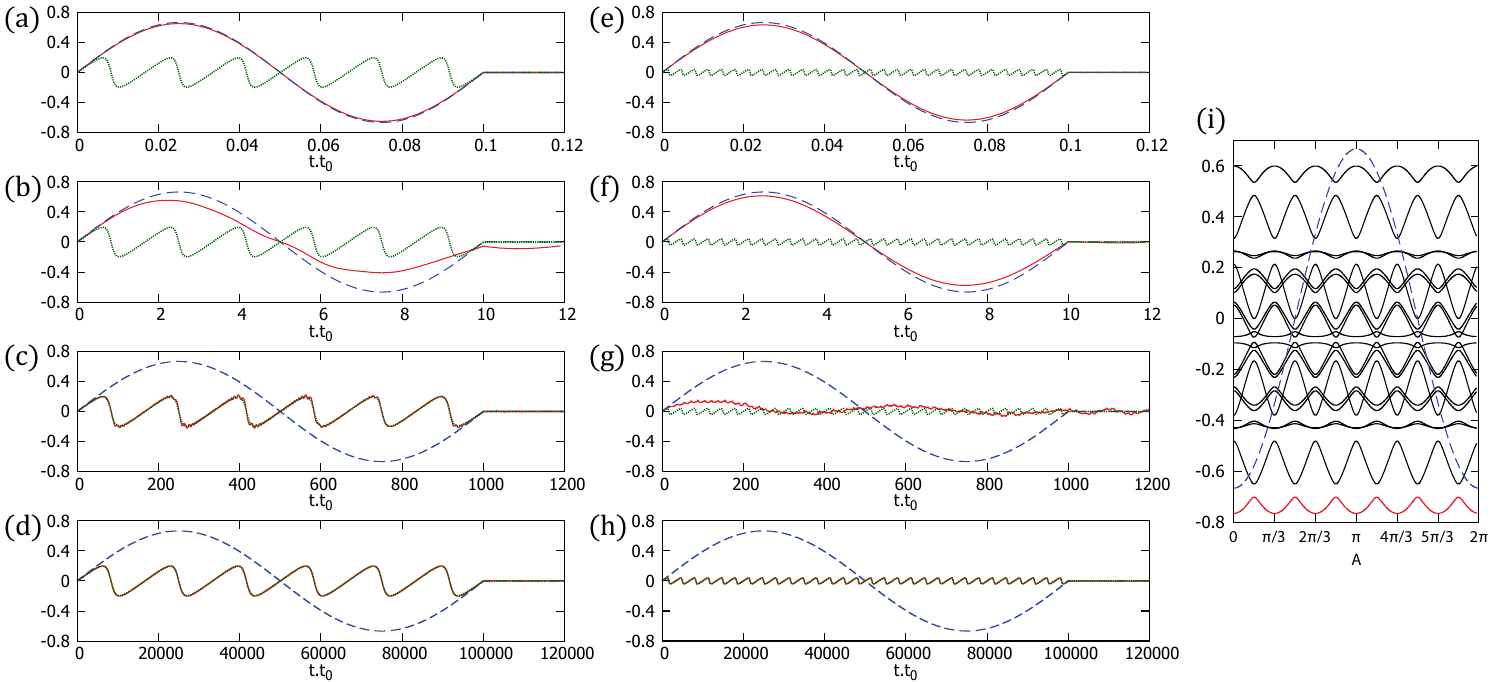}
    \caption{\label{fig2}
    (a--d) [(e--f)] Real-time evolution of the current density $j^L(t)$ (red curve) for $L=6$ [$L=30$] driven by the time-dependent Hamiltonian with a single impurity potential~\eqref{H(t)_w}. The ramp time is set to $T=10^{-1}/t_0, 10/t_0, 10^3/t_0$ and $10^5/t_0$ in (a), (b), (c) and (d) [(e), (f), (g) and (h)], respectively. The total flux $A_0$ is $2\pi$ and the defect energy $w$ is $t_0$. The blue and green dashed curves represent the adiabatic current density for $w=0$ and $w \neq 0$. (i) Many-body energy levels of the Hamiltonian \eqref{H0} with a defect potential \eqref{impuritypotential} for $w=t_0$ and $L=6$. The red curve is the ground state energy density. The blue curve denotes the adiabatic energy spectrum connected to the ground state without flux when $w=0$. The unit in the vertical axis is $t_0$ for all the panels.
    }
    \end{center}
    \end{figure*}

\section{Real-time dynamics}
\label{sec:realtime}

While the optical conductivity and Drude weight have been mostly discussed in the frequency space,
it is also useful to formulate the optical conductivity in terms of real-time response to
the AB flux insertion~\cite{Oshikawa-Drude,OkaAritaAoki2003,IlievskiProsen-DrudeWeights_CMP2013,PhysRevB.102.165137}.
To clarify the physical implication of the divergent behavior of Kohn--Drude weights, we perform a numerical calculation on the tight-binding model considered in the previous section.
We directly simulate the real-time dynamics under a static electric field. Thanks to the simplicity of the model, we can study the real-time dynamics accurately in much detail.

We consider the dynamics driven by the time-dependent Hamiltonian, 
\begin{align}
\hat{H}^{L,\theta(t)}&=-t_0\sum_{x=-L/2+1}^{L/2}(\hat{c}_{x+1}^\dagger e^{-i\theta(t)/L}\hat{c}_x+\mathrm{h.c.})+w \hat{c}^\dagger_0 \hat{c}_0.
\label{H(t)_w}
\end{align}
The systems size $L$ and  the number of electrons is $N_{\mathrm{el}}$ are set to be $4\ell+2$ and $2\ell+1$, respectively.  Here, we use the single potential disorder~\eqref{impuritypotential} as an example. Our results should be independent of the detail of the defect as discussed in Sec.~\ref{sec:twist}. In Appendix~\ref{AppC}, we examine the bond disorder~\eqref{bonddisorder} and indeed obtain essentially the same result. The flux $\theta(t)$ is set to
\begin{align}
    \frac{\theta(t)}{L}=
    \begin{dcases}
    \frac{A_0}{T} t  & (0 \leq t < T)\\
    A_0 & (T \leq t)
    \end{dcases}. \label{theta(t)}
\end{align}
This flux insertion corresponds to the application of the static electric field $E=A_0/T$ within $0\leq t \leq T$. For convenience, we call $T$ the ramp time. The initial state $\ket{\psi_\mathrm{ini}}$ is set to the many-body ground state of $\hat{H}^{L,0}$. For numerical calculation, we discretize the time evolution operator as $\hat{U}(t)\equiv \Delta \hat{U}_{N_t} \cdots \Delta \hat{U}_{2} \Delta \hat{U}_{1}$ where $\Delta \hat{U}_n=\exp[-i\hat{H}^{L,\theta(t_n+\Delta t/2)}\Delta t]$, $\Delta t = t/N_t$, and $t_n=(n-1)\Delta t$. The integer $N_t$ is taken to be large enough to make the result independent of it. The observable we focus on is the time-dependent current density $j^L(t)$ given by the expectation value of the current operator
\begin{align}
    j^L(t) = \langle\hat{U}(t)^\dagger \hat{j}^{L,\theta(t)}\hat{U}(t)\rangle. \label{current}
\end{align}

\begin{figure*}[t]
\begin{center}
\includegraphics[width=11cm]{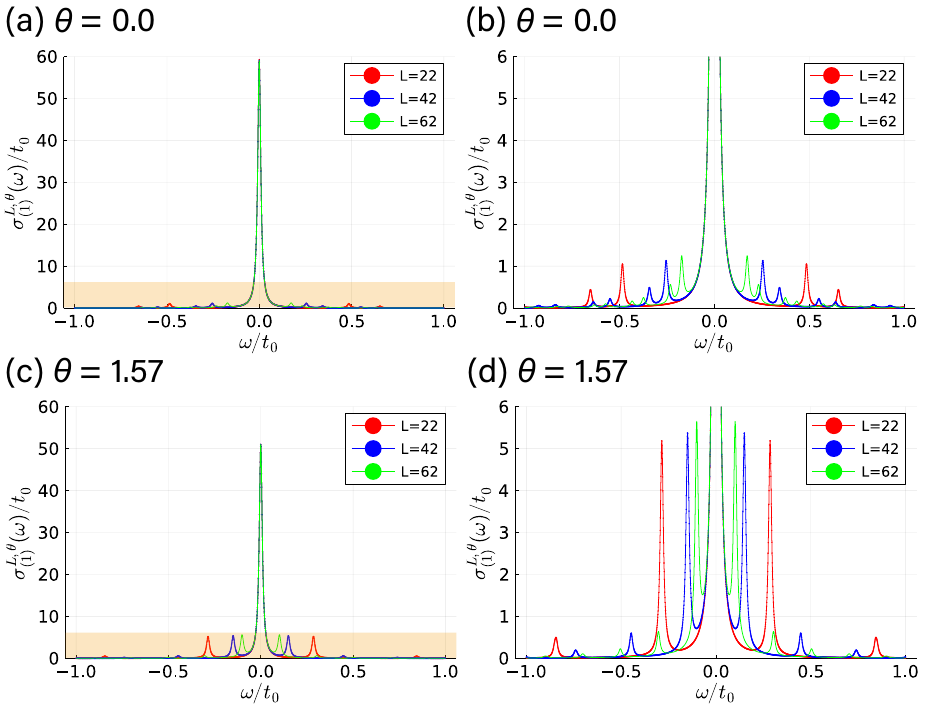}
\caption{
Linear optical conductivity $\sigma_{(1)}^{L, \theta}(\omega)$ for the single-defect model [the Hamiltonian \eqref{H0} with a defect potential \eqref{impuritypotential}] with the different system sizes $L=22, 42, 62$.  We set $w/t_0=1$ and $\eta/t_0=0.01$ in these plots. The panels (a) and (b) are for $\theta=0$ and (c) and (d) are for $\theta=1.57 \sim \pi/2$.  The panels (b) and (d) are the same as  (a) and (c), respectively, but the scale of the vertical axis is expanded.
The yellow shaded region in (a) [(c)] is shown in (b) [(d)].
\label{fig:linear_opt}
}
\end{center}
\end{figure*}

We are interested in the adiabatic transport which is dominated by the Kohn--Drude weights. To this end, we simulate the dynamics for various ramp times. There are two energy scales in this system: the energy gap induced by the defect $\Delta_V\equiv v_Fr_F/L$ in the many-body spectrum [Fig.~\ref{fig2} (i)] and the energy scale of the flux insertion $\Delta_\theta\equiv 2\pi/T$. 
The adiabatic condition is given by $\Delta_\theta\ll\Delta_V$. This is satisfied with the sufficiently large $T$. Under this condition, the current density $j^L(t)$ should be governed by $j_0^{L,\theta}$ in Eq.~\eqref{jGS} which is directly related to the Kohn--Drude weights. On the other hand, $\Delta_\theta\gg\Delta_V$ corresponds to the sudden quench.

The current density $j^L(t)$ is shown in Fig.~\ref{fig2}~(a)-(h). Figs.~2~(a)~and~(e) are near the quench limit. They show the usual Bloch oscillation with the period $T_B=2\pi/E=(2\pi/A_0)T$. The behavior is almost the same as the $r_F=0$ case which is $j^L(t)=(2 t_0/\pi)\sin[\theta(t)/L]$ denoted by the blue dashed curve. For longer ramp time, the current amplitude is suppressed [Figs.~2~(b)~and~(f)] and the profile approaches the different oscillational modes [Figs.~2~(c)~and~(g)]. Finally, it reaches a qualitatively different periodic oscillation profile which has a smaller amplitude and shorter period $T_B^\prime=T_B/L$ [Figs.~2~(d)~and~(h)]. This profile converges to the leading term of the adiabatic current with the defect, i.e., $j^L(t) \sim d_1(\theta(t))/L$. Therefore, this dynamics reaches the adiabatic limit for the given system size $L$. As shown in the previous section, the functional form of the $d_1(\theta)$ implies the divergence of the nonlinear Kohn--Drude weight and this real-time dynamics reflects the divergence. However, the current response is not enhanced with increasing the system size. This shows that the divergence does not necessarily imply a large current response. This observation is consistent with the argument from the analytical results in Sec.~III~E.

\begin{figure*}[t]
\begin{center}
\includegraphics[width=12cm]{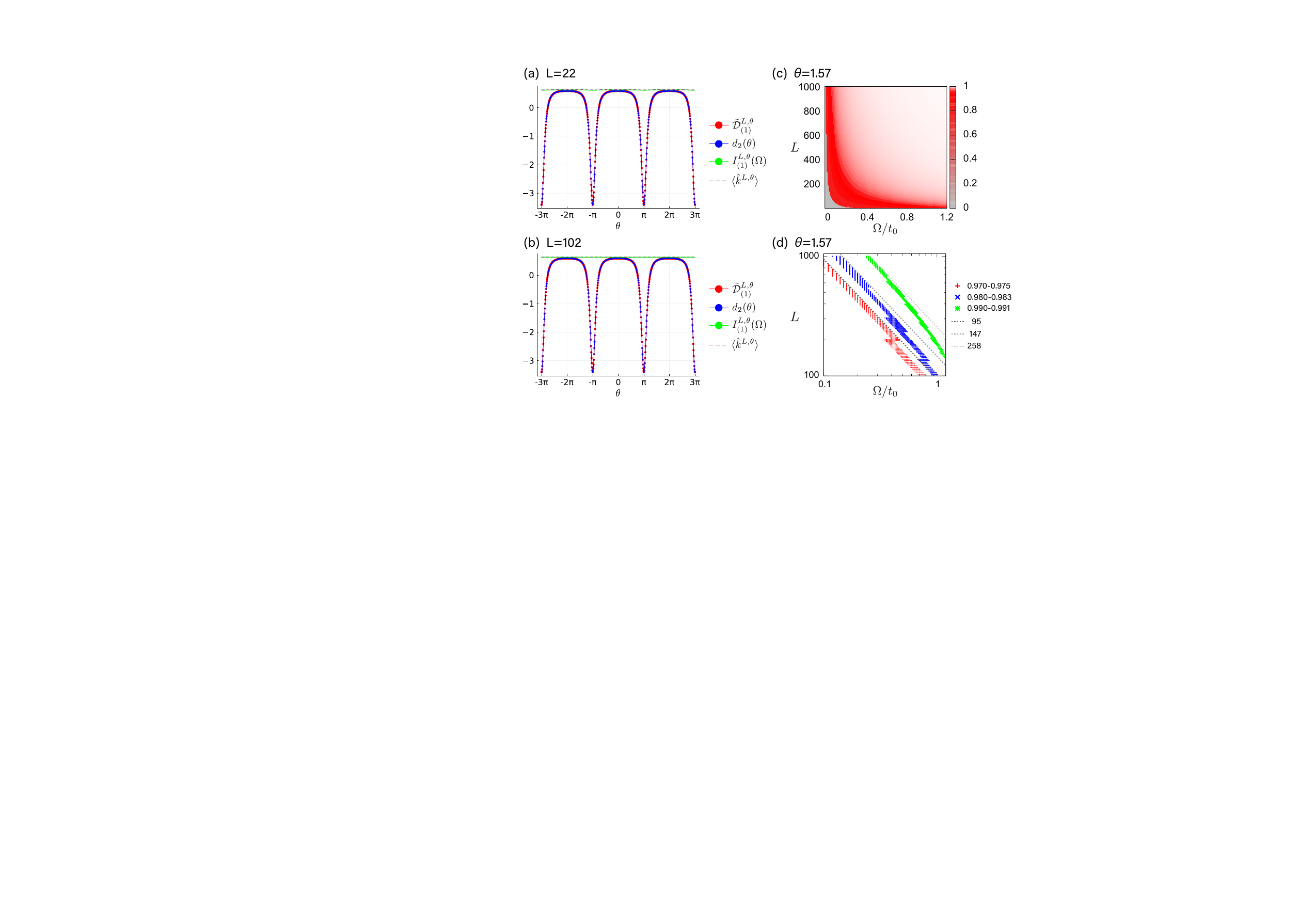}
\caption{
(a, b) Comparison of $\tilde{\mathcal{D}}^{L,\theta}_{(1)}$, $d_2(\theta)$, $I_{(1)}^{L,\theta}(\Omega)$, and $\langle \hat{k}^{L,\theta} \rangle$ for the single-defect model [the Hamiltonian \eqref{H0} with a defect potential \eqref{impuritypotential}]. Here, $\Omega/t_0$ is set to 1 in the panel (a) and 0.5 in the panel (b). The unit in the vertical axis is $t_0$ in both (a) and (b). (c) Color plot of the ratio $r_{(1)}^L(\Omega)$ as a function of $L$ and $\Omega$ for $\theta=1.57 \sim \pi/2$. (d) Regions where $r_{(1)}^L(\Omega)$ takes a value within certain ranges, 0.97-0.975, 0.98-0.983, and 0.99-0.991. Gray dashed lines represent $L=c\Omega^{-1}$ with $c=95, 147$, and $258$. For all the panels, we set $w/t_0=1$.
\label{fig:I1}
}
\end{center}
\end{figure*}

The result of our real-time simulation suggests a way to experimentally measure the nonlinear Kohn--Drude weights through an experiment on the persistent current.
For example, both the flux insertion and the transport measurement have been realized in ultracold atomic systems~\cite{Amico2005, Ryu2007, Cominotti2014, Eckel2014, Lacki2016}. The mesoscopic systems like a metallic ring can also be a realistic platform where the persistent current has been accurately measured~\cite{Bleszynski-Jayich2009, Bluhm2009}. Nonlinear transport in mesoscopic systems is theoretically studied recently~\cite{Kane2022,KawabataNonlinear2021} and  we expect that it provides useful information for the nonlinear Drude weights in finite systems.

Finally, we comment on the crossover-like behavior of the induced current $j^L(t)$, which can be understood from the many-body adiabatic spectrum shown in Fig.~\ref{fig2}~(i). 
In the quench limit, the defect energy scale $\Delta_V$ is much smaller than the flux insertion one $\Delta_\theta$, implying that the defect is irrelevant. In the absence of the defect, the momentum of each electron is conserved and the adiabatic spectrum is $2\pi$-periodic in terms of $\theta/L$ [the blue dashed curve in Fig.~\ref{fig2}~(i)]. This periodicity appears as the Bloch oscillation in the quench limit. Even with the defect, almost perfect non-adiabatic transitions occur at every gap induced by the defect and then the $2\pi$-periodic behavior appears as shown in Fig.~\ref{fig2}~(a)~and~(e). 
On the other side, when $\Delta_V\gg\Delta_\theta$, such non-adiabatic transitions are absent and the time-dependent state keeps sitting on the instantaneous ground state. Such a trajectory is shown in the red curve in Fig.~\ref{fig2}~(i) and this corresponds to the small oscillation with the period $T_B^\prime$ in the adiabatic limit [Figs.~\ref{fig2}~(d)~and~(h)]. In the intermediate scale, non-adiabatic transitions occur in a very complex way and there also appear interferences in the spectrum. Note that a similar crossover behavior and the interference effect appear in the flux insertion in the Hubbard model~\cite{OkaAritaAoki2003, Oka2005QW} and the XXZ model~\cite{Liu2021} in one dimension. This kind of behavior is expected to appear typically in the many-body quantum systems.


\section{Bulk Drude weights}
\label{sec:Dbulk}

As we reviewed in Sec.~\ref{sec:review}, the Kohn--Drude weight
is defined in the adiabatic limit of a finite system.
While it is a perfectly well-defined quantity for the finite-size system, it shows pathological behaviors in the thermodynamic limit:
(i) the twist-angle dependence and (ii) the divergence in the large size limit.
Both of them are critical issues for finding a quantity to characterize the transport property of the bulk.
For this purpose, we need to take the thermodynamic limit first, which is then followed by the zero frequency limit.
While this ``order of limits'' issue has been
discussed~\cite{Oshikawa-Drude,IlievskiProsen-DrudeWeights_CMP2013,Sirker-LesHouches2018,Bertini2021}
for the linear Drude weight, to clarify the issue we call the Drude weight, including nonlinear ones,
defined by taking the thermodynamic limit first as \emph{bulk Drude weight}.
In this section, for the simple single--defect system we demonstrate that, low but non-zero frequency weights in finite size systems contribute to the bulk Drude weight, in which the pathological behaviors disappear.

\subsection{Linear Drude weight}
To see the importance of low-frequency contributions, let us calculate the optical conductivity $\sigma_{(1)}^{L, \theta}(\omega)$ for our single-defect model [Hamiltonian \eqref{H0} with a defect potential \eqref{impuritypotential}]. We assume that the systems size is $L=4\ell+2$ ($\ell\in\mathbb{Z}$) and the particle number is $2\ell+1$. The results for different values of $\theta$ and $L$ are shown in Figs.~\ref{fig:linear_opt}.  In addition to the Drude peak at $\omega=0$, there appear several peaks at finite frequencies. These peaks approach zero frequency with increasing the system size. Indeed, the excitation energy corresponding to these peaks is proportional to $1/L$. These peaks contribute to the low-frequency response in the thermodynamic limit and need to be taken into account to form a well-defined bulk quantity.

Based on this observation, 
we introduce the integral of the real part of the optical conductivity
\begin{align}
I_{(1)}^{L,\theta}(\Omega)&\equiv\frac{1}{\pi}\int_{-\Omega}^{\Omega}d\omega\mathrm{Re}[\sigma_{(1)}^{L,\theta}(\omega)] \notag\\
&= \tilde{\mathcal{D}}_{(1)}^{L,\theta} + 2L \sum_{\substack{n>0, \\\Delta^{L,\theta}_n<\Omega}} \frac{|\langle n^{L,\theta}|\hat{j}^{L,\theta}|0^{L,\theta}\rangle|^2}{\Delta _{n}^{L,\theta}}\label{eq:I1_general}]
\end{align}
as a function of $\Omega>0$. In going to the second line, we used Eq.~\eqref{eq:linear_opt_reg}. The $\Omega \to\infty$ limit reproduces the sum rule in Eq.~\eqref{fsum}
\begin{align}
    \lim_{\Omega \to\infty} I_{(1)}^{L,\theta}(\Omega) =  \langle \hat{k}^{L,\theta}\rangle,
\end{align}
while the $\Omega \to+0$ limit gives the linear Kohn--Drude weight, defined in Eq.~\eqref{eq:linear_KohnDrude}:
\begin{align}
\lim_{\Omega \to+0} I_{(1)}^{L,\theta}(\Omega) =\tilde{\mathcal{D}}_{(1)}^{L,\theta}= \langle\hat{k}^{L,\theta}\rangle - 2L \sum_{n > 0} \frac{|\langle n^{L,\theta}|\hat{j}^{L,\theta}|0^{L,\theta}\rangle|^2}{\Delta_{n}^{L,\theta}}.
\end{align}
In order to take into account finite frequency corrections, we define the linear bulk Drude weight by
\begin{align}
\mathcal{D}_{(1)}^{\mathrm{bulk}}(\theta)\equiv \lim_{\Omega\to+0}\lim_{L\to\infty}I_{(1)}^{L,\theta}(\Omega).
\end{align}
The order of the limit is essential; if we take $\Omega \to+0$ before the $L \to\infty$ limit, we just obtain the Kohn--Drude weight as we stated.

\begin{figure*}[t]
    \begin{center}
    \includegraphics[width=14cm]{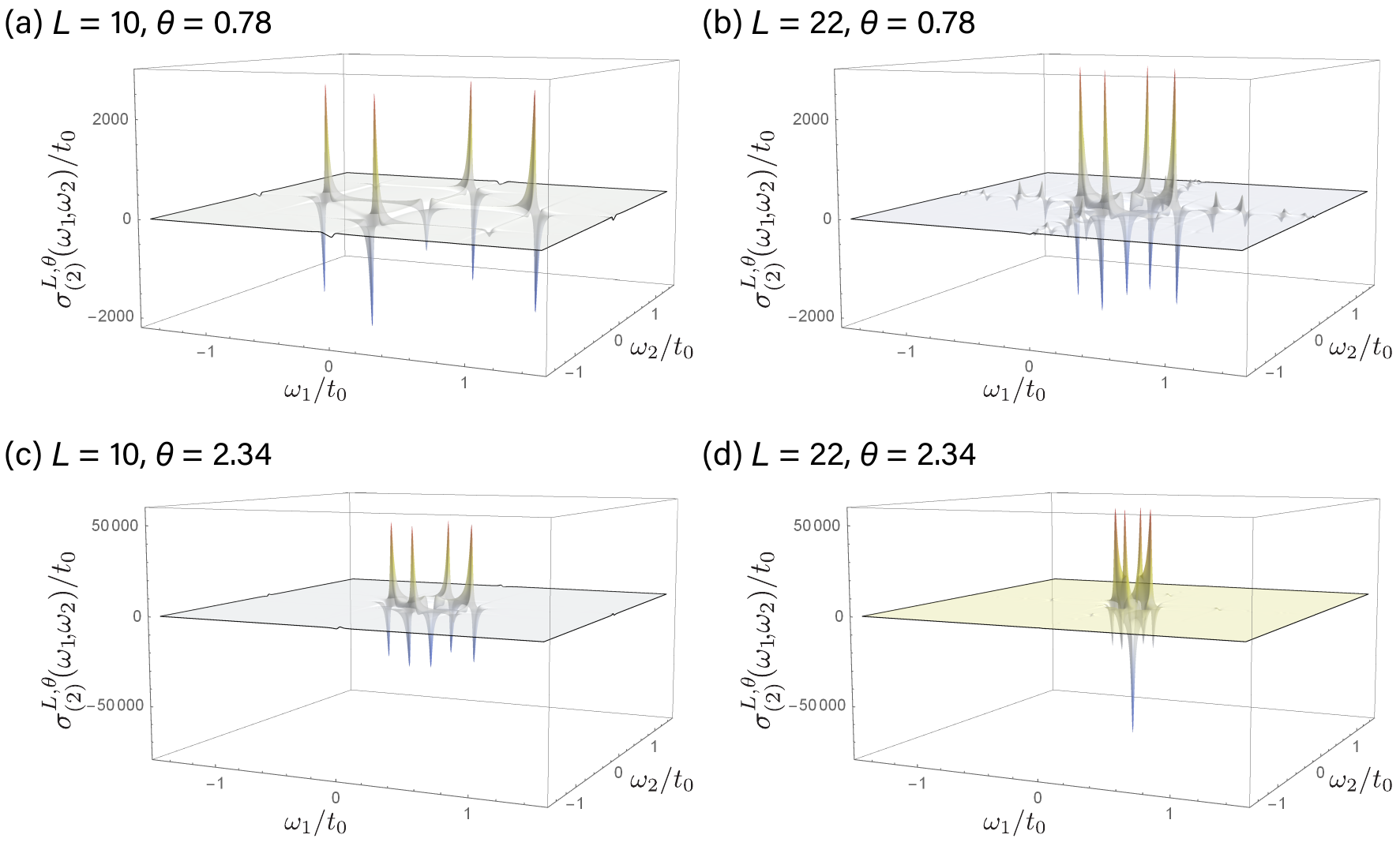}
    \caption{
Second order optical conductivity $\sigma_{(2)}^{L,\theta}(\omega_1, \omega_2)$ for the single-defect model [the Hamiltonian \eqref{H0} with a defect potential \eqref{impuritypotential}] with the different system sizes $L=10, 22$. We set $w/t_0=1$ and $\eta/t_0=0.02$ in these plots. The panels (a) and (b) are for $\theta=0.78 \sim \pi/4$ and (c) and (d) are for $\theta=2.34 \sim 3\pi/4$.
    \label{fig:second_opt}
    }
    \end{center}
    \end{figure*}

We now discuss that  $\mathcal{D}_{(1)}^{\mathrm{bulk}}(\theta)$ behaves as a well-defined bulk quantity by numerically investigating the single impurity model.
We compute $\tilde{\mathcal{D}}_{(1)}^{L,\theta}$ , $d_2(\theta)$, $I_{(1)}^{L,\theta}(\Omega)$ and $\langle \hat{k}^{L,\theta}\rangle$ using Eq.~\eqref{eq:DKohn1_free}, Eq.~\eqref{d2}, Eq.~\eqref{eq:I1_free}, and Eq.~\eqref{eq:linear_opt_free_1}, respectively, and compare them in Figs.~\ref{fig:I1}~(a)~and~(b). As we can see, the twist-angle dependence that appears in $\tilde{\mathcal{D}}_{(1)}^{L,\theta}$ almost vanishes in $I_{(1)}^{L,\theta}(\Omega)$. We find that $I_{(1)}^{L,\theta}(\Omega)$ with $\Omega=t_0$ is almost identical to $\langle \hat{k}^{L,\theta}\rangle$, suggesting that the second term in Eq.~\eqref{eq:linear_KohnDrude} is nearly canceled by the second term in Eq.~\eqref{eq:I1_general}. 
To see this cancellation more clearly, we calculate the ratio
\begin{align}
r_{(1)}^L(\Omega)&\equiv\frac{I_{(1)}^{L,\theta}(\Omega)-\tilde{\mathcal{D}}_{(1)}^{L,\theta}}{\langle \hat{k}^{L,\theta}\rangle-\tilde{\mathcal{D}}_{(1)}^{L,\theta}}\notag\\
&=\frac{\sum_{n>0,\Delta^{L,\theta}_n<\Omega} |\langle n^{L,\theta}|\hat{j}^{L,\theta}|0^{L,\theta}\rangle|^2/\Delta_{n}^{L,\theta}}{\sum_{n > 0} |\langle n^{L,\theta}|\hat{j}^{L,\theta}|0^{L,\theta}\rangle|^2/\Delta_{n}^{L,\theta}},\label{r1}
\end{align}
which takes value in the range $[0,1]$ and becomes one when $\Omega$ is sufficiently large so that the cancellation is perfect. The ratio $r_{(1)}^L(\Omega)$ is plotted for different values of $L$ and $\Omega$ in Fig.~\ref{fig:I1}~(c). We also plot the regions where $r_{(1)}^L(\Omega)$ takes the values near one in Fig.~\ref{fig:I1}~(d), which shows that the contour for $r_{(1)}^L(\Omega_0(L)) \sim 1$ behaves like $\Omega_0(L)\propto L^{-1}$. Based on these plots, we see that $r_{(1)}^L(\Omega)$ converges to one in the $\Omega \to+0$ limit \emph{after} the $L \to\infty$ limit.  We conclude that $\mathcal{D}_{(1)}^{\mathrm{bulk}}(\theta) = \lim_{L \to\infty} \langle\hat{k}^{L}(\theta)\rangle=2t_0/\pi$ and does not depend on $\theta$ in this model.

Finally, we remark on the relation to the previous works on
the linear Drude weights in the open boundary condition (OBC). It was shown that the Kohn--Drude weight vanishes under the OBC in the systems where the Kohn--Drude weight is non-zero under the PBC~\cite{RigolShastry-Drude,BellomiaResta}. These studies have shown that this inconsistency can be resolved by considering the $1/L$-excitations in the OBC case. Similarly to our study, the peaks move to zero frequency with increasing the system size and form a part of Drude peak in the thermodynamic limit. The OBC result is found to be consistent with the PBC one once these low frequency peaks are included. Since the OBC corresponds to the infinitely strong potential barrier, i.e., a potential defect \eqref{impuritypotential} with $w/t_0\to \infty$,
their results can be regarded as an extreme case of our results although for a more nontrivial model with interaction. On the other hand, our analysis on the simple model can be further extended to nonlinear Drude weights, as we will discuss in the following subsection.

\subsection{Second-order Drude weight}

Let us perform the same analysis for the second-order optical conductivity. We plot  $\sigma_{(2)}^{L, \theta}(\omega_1,\omega_2)$ in Fig.~\ref{fig:second_opt}. There appear finite frequency peaks in addition to the Drude peak [$(\omega_1, \omega_2)=(0, 0)$], and they approach the Drude peak with increasing the system size. This behavior is similar to the linear order and thus suggests that the bulk Drude weight can be defined in the same way.  Namely, we introduce the integral of the optical conductivity
\begin{align}
    I_{(2)}^{L,\theta}(\Omega)
\equiv\frac{1}{\pi^2}\int_{-\Omega}^{\Omega}d\omega_1\int_{-\Omega}^{\Omega}d\omega_2\sigma_{(2)}^{L,\theta}(\omega_1,\omega_2)
\end{align}
and define second-order bulk Drude weight by
\begin{align}
\mathcal{D}_{(2)}^{\mathrm{bulk}}\equiv \lim_{\Omega\to+0}\lim_{L\to\infty}I_{(2)}^{L,\theta}(\Omega).\label{defD2phys}
\end{align}

\begin{figure*}[t]
    \begin{center}
    \includegraphics[width=16cm]{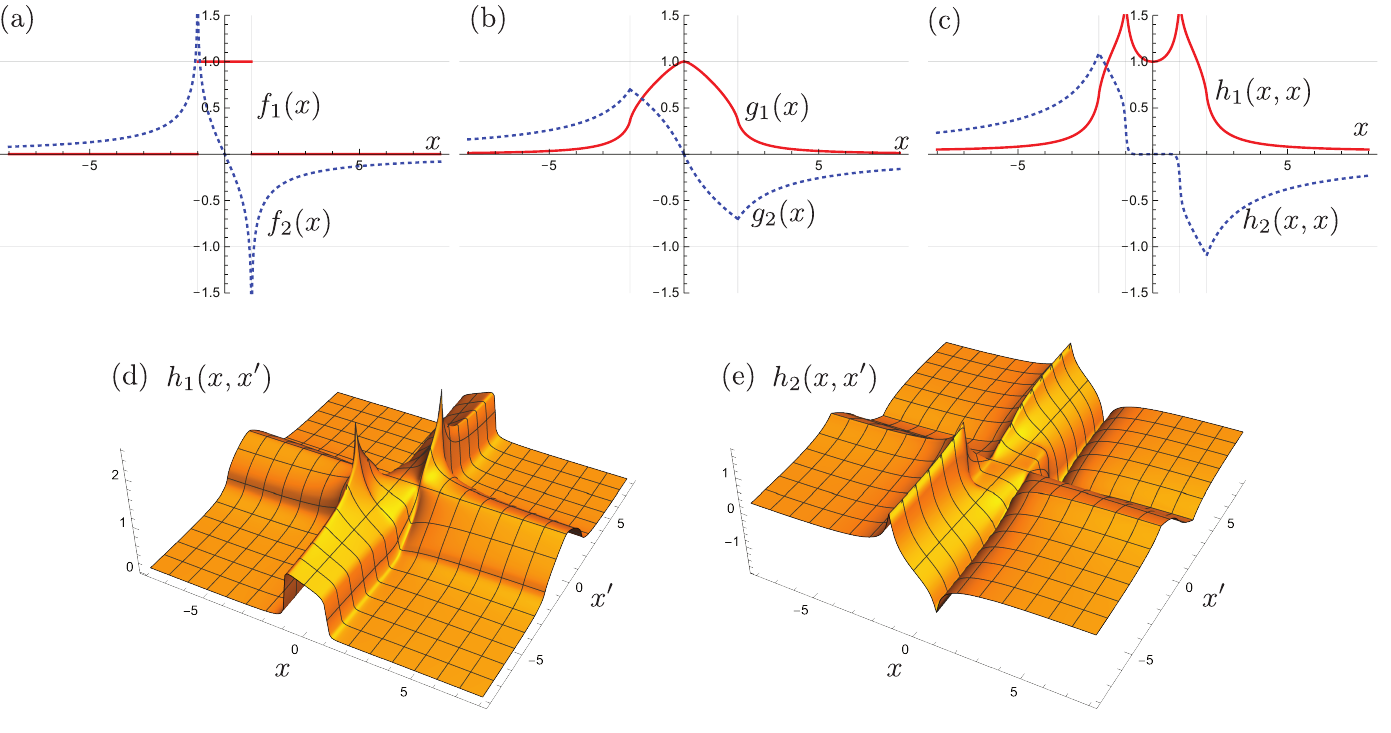}
    \caption{Plot of $f_{1,2}(x)$, $g_{1,2}(x)$, and $h_{1,2}(x,x')$. The function $h_{1,2}(x,x')$ is computed numerically for $\eta/t_0=0.005$. \label{fig:functions}
    }
    \end{center}
\end{figure*}

Let us investigate these quantities in more details. We rewrite $I_{(2)}^{L,\theta}(\Omega)$ using Eqs.~\eqref{eq:second_optcond}-\eqref{eq:second_opt_phi2}:
\begin{widetext}
\begin{align}
I_{(2)}^{L,\theta}(\Omega)
&=\tilde{\mathcal{D}}_{(2)}^{L,\theta}
+2L\sum_{n>0}\frac{\mathrm{Re}[\langle 0^{L,\theta}|\hat{j}^{L,\theta}|n^{L,\theta}\rangle\langle n^{L,\theta}|\hat{k}^{L,\theta}|0^{L,\theta}\rangle]}{\Delta_{n}^{L,\theta}}\left(g_1(\tfrac{\Delta_{n}^{L,\theta}}{\Omega})+2f_1(\tfrac{\Delta_{n}^{L,\theta}}{\Omega})\right)\notag\\
&\quad-2L\sum_{n>0}\frac{\mathrm{Im}[\langle 0^{L,\theta}|\hat{j}^{L,\theta}|n^{L,\theta}\rangle\langle n^{L,\theta}|\hat{k}^{L,\theta}|0^{L,\theta}\rangle]}{\Delta_{n}^{L,\theta}}\left(g_2(\tfrac{\Delta_{n}^{L,\theta}}{\Omega})-2f_2(\tfrac{\Delta_{n}^{L,\theta}}{\Omega})\right)\notag\\
&\quad-2L^2\sum_{m,l>0}\frac{\mathrm{Re}[\langle 0^{L,\theta}|\hat{j}^{L,\theta}|m^{L,\theta}\rangle\langle m^{L,\theta}|\hat{j}^{L,\theta}|l^{L,\theta}\rangle\langle l^{L,\theta}|\hat{j}^{L,\theta}|0^{L,\theta}\rangle]}{\Delta_{m}^{L,\theta}\Delta_{l}^{L,\theta}}\notag\\
&\quad\quad\quad\quad\quad\quad\quad\quad\quad\times\left[2h_1(\tfrac{\Delta_{l}^{L,\theta}}{\Omega},\tfrac{\Delta_{m}^{L,\theta}}{\Omega})+f_1(\tfrac{\Delta_{l}^{L,\theta}}{\Omega})+f_1(\tfrac{\Delta_{m}^{L,\theta}}{\Omega})-f_1(\tfrac{\Delta_{m}^{L,\theta}}{\Omega})f_1(\tfrac{\Delta_{l}^{L,\theta}}{\Omega})-f_2(\tfrac{\Delta_{m}^{L,\theta}}{\Omega})f_2(\tfrac{\Delta_{l}^{L,\theta}}{\Omega})\right]
\notag\\
&\quad+2L^2\sum_{m,l>0}\frac{\mathrm{Im}[\langle 0^{L,\theta}|\hat{j}^{L,\theta}|m^{L,\theta}\rangle\langle m^{L,\theta}|\hat{j}^{L,\theta}|l^{L,\theta}\rangle\langle l^{L,\theta}|\hat{j}^{L,\theta}|0^{L,\theta}\rangle]}{\Delta_{m}^{L,\theta}\Delta_{l}^{L,\theta}}\notag\\
&\quad\quad\quad\quad\quad\quad\quad\quad\quad\times\left[2h_2(\tfrac{\Delta_{l}^{L,\theta}}{\Omega},\tfrac{\Delta_{m}^{L,\theta}}{\Omega})+f_2(\tfrac{\Delta_{l}^{L,\theta}}{\Omega})-f_2(\tfrac{\Delta_{m}^{L,\theta}}{\Omega})-f_1(\tfrac{\Delta_{m}^{L,\theta}}{\Omega})f_2(\tfrac{\Delta_{l}^{L,\theta}}{\Omega})+f_1(\tfrac{\Delta_{l}^{L,\theta}}{\Omega})f_2(\tfrac{\Delta_{m}^{L,\theta}}{\Omega})\right]\notag\\
&\quad+2L^2\sum_{n>0}\frac{\langle \hat{j}^{L,\theta}\rangle\langle 0^{L,\theta}|\hat{j}^{L,\theta}|n^{L,\theta}\rangle\langle n^{L,\theta}|\hat{j}^{L,\theta}|0^{L,\theta}\rangle}{\Delta_{n}^{L,\theta}\Delta_{n}^{L,\theta}}\left[2h_1(\tfrac{\Delta_{n}^{L,\theta}}{\Omega},\tfrac{\Delta_{n}^{L,\theta}}{\Omega})+2f_1(\tfrac{\Delta_{n}^{L,\theta}}{\Omega})-f_1(\tfrac{\Delta_{n}^{L,\theta}}{\Omega})^2-f_2(\tfrac{\Delta_{n}^{L,\theta}}{\Omega})^2\right],
\end{align}
\end{widetext} 
where $\tilde{\mathcal{D}}_{(2)}^{L,\theta}$ is given in Eq.~\eqref{D_2} and functions $f_{1,2}(x)$, $g_{1,2}(x)$, and $h_{1,2}(x,x')$ are defined by
\begin{align}
F(x)&\equiv\frac{-1+f_1(x)+if_2(x)}{x}\notag\\
&\equiv\frac{1}{\pi}\int_{-1}^{1}dy\frac{i}{y+i\eta'}\frac{1}{y-x+i\eta'}\notag\\
&=\frac{\Omega}{\pi}\int_{-\Omega}^{\Omega}d\omega\frac{i}{\omega+i\eta}\frac{1}{\omega-\Delta+i\eta},
\end{align}
\begin{align}
G(x)&\equiv\frac{-1+g_1(x)+ig_2(x)}{x}\notag\\
&\equiv\frac{1}{\pi}\int_{-1}^{1}dy\frac{i}{y+i\eta'}F(x-y)\notag\\
&=\frac{\Omega}{\pi^2}\int_{-\Omega}^{\Omega}d\omega_1\int_{-\Omega}^{\Omega}d\omega_2\frac{i}{\omega_1+i\eta}\frac{i}{\omega_2+i\eta}\notag\\
&\quad\quad\quad\quad\quad\quad\quad\quad\quad\times\frac{1}{\omega_1+\omega_2-\Delta+i\eta},
\end{align}
and
\begin{align}
H(x,x')&\equiv\frac{-1+h_1(x,x')+ih_2(x,x')}{xx'}\notag\\
&\equiv-\frac{1}{\pi}\int_{-1}^{1}dy\frac{i}{y+i\eta'}\frac{1}{y-x+i\eta'}F(x'-y)\notag\\
&=-\frac{\Omega^2}{\pi^2}\int_{-\Omega}^{\Omega}d\omega_1\int_{-\Omega}^{\Omega}d\omega_2\frac{i}{\omega_1+i\eta}\frac{i}{\omega_2+i\eta}\notag\\
&\quad\times\frac{1}{(\omega_1-\Delta+i\eta)(\omega_1+\omega_2-\Delta'+2i\eta)}.
\end{align}
In these expressions, we wrote $x=\Delta/\Omega$,  $x'=\Delta'/\Omega$, $y=\omega/\Omega$, and $\eta'=\eta/\Omega$. We find the following analytic expressions for $f_{1,2}(x)$ and $g_{1,2}(x)$:
\begin{align}
f_1(x)&\equiv \theta(1-|x|)=f_1(-x),\\
f_2(x)&\equiv\frac{1}{\pi}\log\frac{|1-x|}{|1+x|}=-f_2(-x),
\end{align}
\begin{align}
g_1(x)&\equiv \frac{2}{\pi^2}\text{Re}\Big[\text{Li}_2(1+x)-\text{Li}_2(-1-x)\notag\\
&\quad\quad\quad\quad\quad+\text{Li}_2(1-x)-\text{Li}_2(-1+x)\Big]\notag\\
&=g_1(-x),
\end{align}
\begin{align}
g_2(x)&\equiv\frac{2}{\pi^2}\text{sign}(x)\text{Im}\Big[\text{Li}_2(1+x)-\text{Li}_2(-1-x)\notag\\
&\quad\quad\quad\quad\quad\quad\quad\quad+\text{Li}_2(1-x)-\text{Li}_2(-1+x)\Big]\notag\\
&=-g_2(-x),
\end{align}
where $\text{Li}_s(z)\equiv \sum_{n=1}^\infty z^n/n^s$ is the polylogarithm function. While we do not have an analytic expression for $H(x,x')$, it can be numerically calculated using $F(x)$. 
These functions are plotted in Fig.~\ref{fig:functions}.

\begin{figure*}[t]
\begin{center}
\includegraphics[width=12cm]{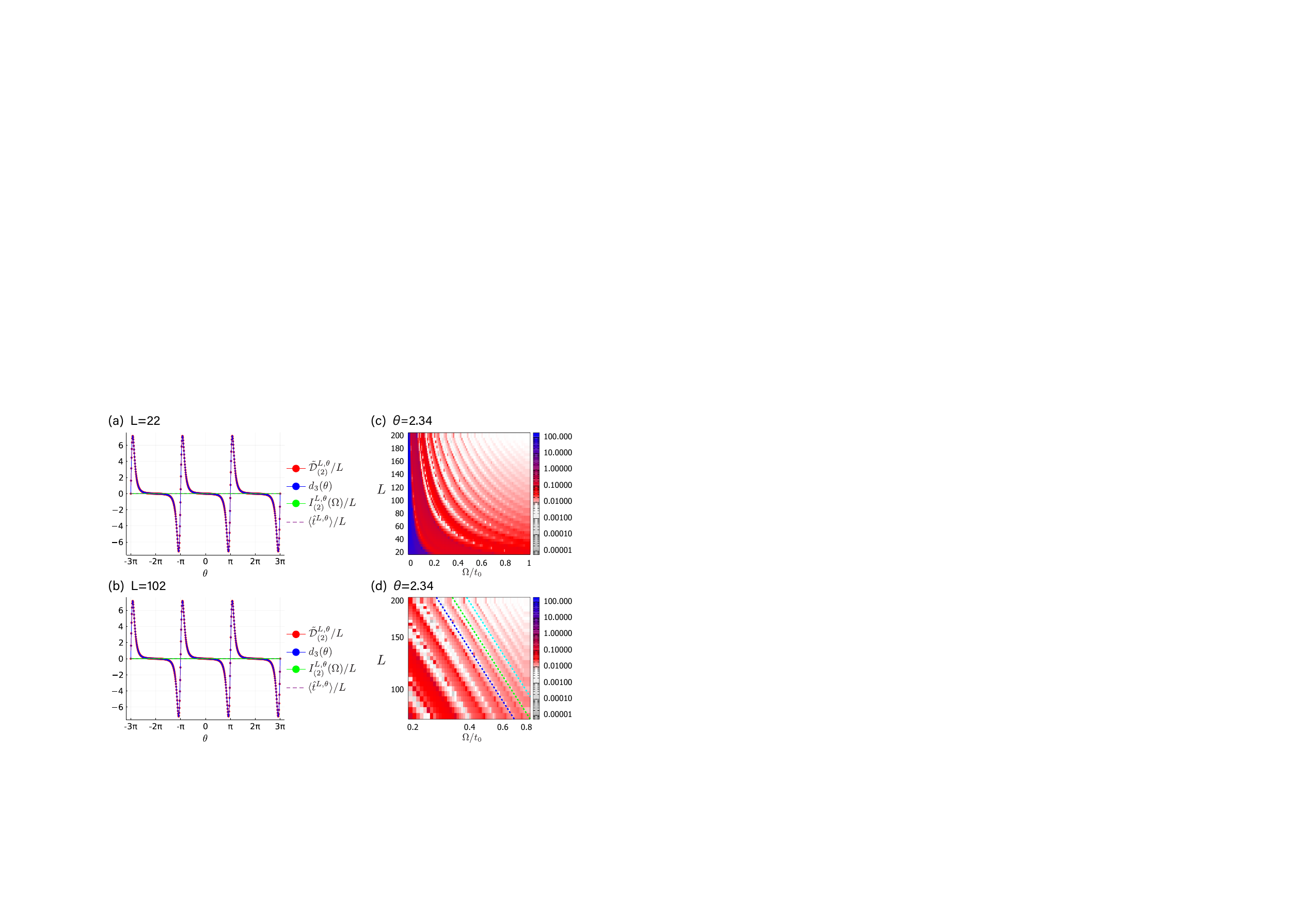}
\caption{
(a, b) Comparison of $\tilde{\mathcal{D}}^{L,\theta}_{(2)}$, $d_3(\theta)$, $I_{(2)}^{L,\theta}(\Omega)$, and $\langle \hat{t}^{L,\theta} \rangle$ for the single-defect model [the Hamiltonian \eqref{H0} with a defect potential \eqref{impuritypotential}]. Here, $\Omega/t_0$ is set to 1 in the panel (a) and 0.5 in the panel (b). Note that the plotted values except for $d_3(\theta)$ are divided by $L$. (c) Color plot of the ratio $I_2^{L,\theta}(\Omega)$ as a function of $L$ and $\Omega$ for $\theta=2.34 \sim 3\pi/4$. (d) is the expanded panel (c) with logarithmic scale. We also plot $L=c \Omega^{-1}$ ($c=56, 68$, and $81$) as a guide by colored dashed lines. We set $\eta/t_0=0.005$ in these plots. For all the panels, we set $w/t_0=1$.
\label{fig:I2}
}
\end{center}
\end{figure*}

Using these expressions, let us study the behaviors of $I^{L,\theta}_{(2)}(\Omega)$. 
For example, the $\Omega\to\infty$ limit gives the generalized frequency sum rule~\cite{PhysRevB.102.165137}
\begin{align}
\lim_{\Omega\to\infty}I^{L,\theta}_{(2)}(\Omega)=\langle \hat{t}^{L,\theta} \rangle,
\end{align}
since $f_{1}(x)$, $g_{1}(x)$, and $h_{1}(x,x')$ become 1 and $f_{2}(x)$, $g_{2}(x)$, and $h_{2}(x,x')$ become 0 in the $|x|, |x'|\rightarrow0$ limit. 
On the other hand, the $\Omega\to +0$ limit gives the Kohn--Drude weight 
\begin{align}
\lim_{\Omega\to+0}I^{L,\theta}_{(2)}(\Omega)=\tilde{\mathcal{D}}_{(2)}^{L,\theta},
\end{align}
since $f_{1,2}(x)$, $g_{1,2}(x)$, and $h_{1,2}(x,x')$ all become $0$ in the $|x|, |x'|\rightarrow\infty$ limit.

To demonstrate the well-definedness of the bulk Drude weight $\mathcal{D}_{(2)}^{\mathrm{bulk}}$, we compute $\tilde{\mathcal{D}}_{(2)}^{L,\theta}$, $d_3(\theta)$, and $I_{(2)}^{L,\theta}(\Omega)$ using Eq.~\eqref{eq:DKohn2_free}, Eq.~\eqref{d3}, and Eq.~\eqref{eq:I2_free}, respectively. We plot them in Figs.~\ref{fig:I2}~(a)~and~(b). As shown in these figures, $I_{(2)}^{L,\theta}(\Omega)$ is almost zero for any twist angle $\theta$, implying that the bulk Drude weight $\mathcal{D}_{(2)}^{\mathrm{bulk}}(\theta)$ vanishes. This is expected because the second-order response is prohibited by the spatial-inversion symmetry when $\theta=0$ and $\mathcal{D}_{(2)}^{\mathrm{bulk}}(\theta)$ should not depend on $\theta$.
To confirm this, we calculate $|I_{(2)}^{L,\theta}(\Omega)|$ for different $L$ and $\Omega$. The results are shown in Fig.~\ref{fig:I2}~(c). This figure is qualitatively similar to the linear-order one [Fig.~\ref{fig:I1}~(c)] while there is an additional oscillation not existing in the linear one. Thus, as discussed in the previous subsection, this suggests that $I_{(2)}^{L,\theta}(\Omega)$ becomes zero in the $\Omega \to+0$ limit after the $L \to\infty$ limit. We also show the expanded version of this figure with logarithmic scale in Fig.~\ref{fig:I2}~(d). The region where $|I_{(2)}^{L,\theta}(\Omega)|$ approaches zero has the shape like the line $L \propto \Omega^{-1}$ that is plotted as a guide in Fig.~\ref{fig:I2}~(d). Thus we conclude that  $\mathcal{D}_{(2)}^{\mathrm{bulk}}(\theta)=\lim_{\Omega \to+0} \lim_{L \to\infty} I_{(2)}^{L,\theta}(\Omega)=0$ in this model.

\section{Discussion}
\label{sec:Discussion}
We have shown that the pathological behaviors of the Kohn--Drude weights in the presence of a single defect are resolved by considering the bulk Drude weight that contains additional contributions from the regular part. While we have only provided explicit calculations for the linear and the second-order responses, we expect that similar phenomena occur in the higher orders. We define the $N$-th order bulk Drude weight as
\begin{align}
    \mathcal{D}^\mathrm{\mathrm{bulk}}_{(N)} &\equiv \lim_{\Omega \to+0} \lim_{L \to\infty} I^{L,\theta}_{(N)}(\Omega),\label{eq:Dphys_gen} \\
    I^{L,\theta}_{(N)}(\Omega) &\equiv \frac{1}{\pi^{N}}  \int_{-\Omega}^{\Omega}\frac{d\omega_1}{2\pi} \cdots \int_{-\Omega}^{\Omega}\frac{d\omega_N}{2\pi} \sigma^{L,\theta}_{(N)}(\omega_1, \cdots, \omega_N),
\end{align}
where $\sigma_{(N)}^{L,\theta}(\omega_1, \cdots, \omega_N)$ is the $N$-th order optical conductivity with the system size $L$ and the twist angle $\theta$. $\mathcal{D}^{\mathrm{bulk}}_{(N)}$ should work as a well-defined bulk quantity, as we have shown for $N=1, 2$. We leave the rigorous proof as a future work.

Although we have discussed only the single-defect model in this work, our conclusion should be applicable more generally 
because the essential ingredient for our argument was merely the presence of $1/L$-excitations. For example, the spin-1/2 XXZ chain is known to show the divergence of the nonlinear Drude weight~\cite{PhysRevB.102.165137,Tanikawa2021,Fava,Tanikawa2021fine} and the discontinuous change with respect to the anisotropy~\cite{Liu2021,Tanikawa2021fine}. Since the XXZ spin chain also shows the $1/L$-excitation in the gapless regime, our approach is expected to be applicable to this case.

Here we point out an important universal relation among
the \emph{linear} Kohn--Drude and bulk Drude weights, and $f$--sum:
\begin{equation}
    \lim_{L\to\infty}\tilde{\mathcal{D}}_{(1)}^{L,\theta}
    \leq
    \mathcal{D}_{(1)}^{\mathrm{bulk}}
    \leq
    \lim_{L\to\infty}\frac{1}{\pi}\int_{-\infty}^{\infty}d\omega\sigma_{(1)}^{L,\theta}(\omega).
\label{eq:Kohn_bulk_fsum}
\end{equation}  
This is a simple consequence of the non-negativity of the optical conductivity
\begin{equation}
\sigma_{(1)}^{L,\theta}(\omega)\geq 0,
\label{eq:sigma_geq_0}
\end{equation}
which is apparent from Eq.~\eqref{eq:linear_opt_reg}.
That is, while the Kohn--Drude weight only reflects the delta-function peak strictly at zero frequency in
finite-size systems, the bulk Drude weight also includes the conductivities at nonzero but small frequencies in finite-size
systems. Finally, the $f$-sum contains the conductivities over the entire frequency range.
Because of the non-negativity, these quantities must be in the ascending order.
Therefore, even though the (linear) Kohn--Drude weight is defined in a
``wrong'' order for characterizing the bulk property,
it can still serve as a lower bound for the bulk Drude weight.
The Mazur bound for the Kohn--Drude weight, which is based only
on charges exactly conserved in finite size systems, also works as a lower bound for the bulk Drude weight as well
(although it is generally weaker than the Mazur bound for the bulk Drude weight including the quasi--conserved charges). Although we mostly focus on the zero temperature limit in this paper, the non-negativity~\eqref{eq:sigma_geq_0}
and thus the inequality~\eqref{eq:Kohn_bulk_fsum} are \emph{valid for any temperature}.

On the other hand, the nonlinear AC conductivities are in general not subject to the non-negativity constraint,
and become indeed negative at some frequencies even in simple tight-binding models~\cite{WatanabeLiuOshikawa}.
This implies that, the naive \emph{nonlinear} generalization of Eq.~\eqref{eq:Kohn_bulk_fsum}
on the Kohn--Drude weight, the bulk Drude weight, and $f$-sum do \emph{not} generally hold.
The general relation among these three quantities for nonlinear conductivities will be discussed in
a separate publication~\cite{Takasan_future}.

For the special case of the single-band tight-binding model we have studied in this paper,
the relation~\eqref{eq:Kohn_bulk_fsum} can be further simplified. 
First let us consider the periodic boundary condition without a defect.
The single-particle energy eigenstates are then given by plane waves.
The insertion of the AB flux effectively shifts the momentum of the each plane wave state.
In the single-band model, there is a unique energy eigenstate for each quantized momentum.
Therefore, the final state after the AB flux insertion is completely determined by the final amount of 
the AB flux, and does not depend on the schedule (including the speed) of the flux insertion.
This means that the generic AB flux insertion is equivalent to the adiabatic limit, implying that
$\sigma_{(1)}^{L,\theta}(\omega)$ consists only of the delta-function peak at $\omega=0$. Therefore, in the single-band tight-binding model with the periodic boundary condition, 
\begin{equation}
    \lim_{L\to\infty}\tilde{\mathcal{D}}_{(1)}^{L,\theta}
    =
    \mathcal{D}_{(1)}^{\mathrm{bulk}}
    =
    \lim_{L\to\infty}\frac{1}{\pi}\int_{-\infty}^{\infty}d\omega\sigma_{(1)}^{L,\theta}(\omega) .
\label{eq:Kohn_bulk_fsum_eq}
\end{equation}   
Even for general boundary conditions, the second equality should hold, since the bulk Drude weight
must be independent of the boundary condition, although the first equality is reduced to the
inequality as in Eq.~\eqref{eq:Kohn_bulk_fsum}.

The same argument also applies to nonlinear conductivities.
For a single-band tight-binding model with the periodic boundary condition,
the generic AB flux insertion is equivalent to the adiabatic limit
at any order.
Thus the straightforward generalization of Eq.~\eqref{eq:Kohn_bulk_fsum_eq}
holds for the nonlinear conductivity at every order, under the periodic boundary condition.
The second equality between the bulk Drude weight and the $f$-sum still holds under general boundary conditions,
while the first equality is lost in general.
We stress that this is a special property of the single-band model.
In a multi-band model, even if the system is noninteracting, a non-adiabatic AB flux insertion generally
causes inter-band transitions and is inequivalent to the adiabatic AB flux insertion.

\section{Summary and Outlook}
\label{sec:Summary}
In this work, we clarified the dependence of the ground state energy on the twisted boundary condition in 1D systems in general. We derived a general upper bound \eqref{newbound} of the adiabatic current density in terms of the frequency sum of the optical conductivity, which may be regarded as a refined version of the Bloch theorem. As an illuminating toy model, we discussed a single-band tight-binding model in the presence of a single defect.
Our study on the simple model illustrates the importance of
the order of limits~\protect\cite{Oshikawa-Drude,IlievskiProsen-DrudeWeights_CMP2013,Sirker-LesHouches2018,Bertini2021}
in defining the Drude weight,
especially the nonlinear ones.
In order to clarify the issue, we call the thermodynamic limit of the coefficient of the zero-frequency
delta function in the AC conductivity of finite-size systems as Kohn--Drude weight, whereas bulk Drude weight
is defined by taking the thermodynamic limit before the zero-frequency limit.

We found that the linear Kohn--Drude weight $\mathcal{D}_{(1)}^{L,\theta}$ in the large-$L$ limit depends nontrivially on the twist angle $\theta$ due to the presence of the defect. 
We also found that $N$-th order  Kohn--Drude weights $\tilde{\mathcal{D}}_{(N)}^{L,\theta}$ ($N\geq2$) in our model exhibits a strong divergence proportional to $L^{N-1}$ in the large-$L$ limit. Then, we studied the physical implication of the divergence through the direct numerical simulation which is relatively easy since this model is noninteracting. 
Furthermore, we showed how the finite but small frequency components in the finite-size systems contribute   
to the bulk Drude weight. There are low-energy (which scales as $1/L$ for the system size $L$)
excitations, which appear in the
AC conductivity at finite ($O(1/L)$) frequencies in the finite-size systems.
By taking the thermodynamic limit first, these components are merged into the delta-function peak at zero frequency,
thus contributing to the bulk Drude weight.
This also eliminates various pathological behaviors, especially the divergence, of the Kohn--Drude weight
which should be absent in the bulk.

The Kohn--Drude weight, on the other hand, is a perfectly well-defined quantity for finite-size systems. It could be measured experimentally in, for example, cold atoms placed on a ring~\cite{Amico2005, Ryu2007, Cominotti2014, Eckel2014, Lacki2016} or electrons in a metallic ring~\cite{Bleszynski-Jayich2009, Bluhm2009}.
Furthermore, sometimes the thermodynamic limit and the zero-frequency limit may be exchangeable.
When this is the case, Kohn--Drude weight and bulk Drude weight are identical. Previous studies~\cite{IlievskiProsen-DrudeWeights_CMP2013, Bertini2021} imply that this is the case of the linear Drude weight in the $S=1/2$ XXZ chain at zero temperature.
Furthermore, as we have pointed out in Sec.~\ref{sec:Discussion}, the two limits are identical
at all orders of linear or nonlinear Drude weights,
for the single-band tight-binding model with the periodic boundary condition without a defect.
The conditions for the nonlinear Kohn--Drude and bulk Drude weights to be identical in more general systems
are left for future investigations.

Our results open various directions for future studies. While we only studied a single defect, it is an attractive problem to study the multi-defect (disordered) model, which should be relevant to the physics of Anderson localization. The effect of interaction in the single-defect model is also interesting because it has been already known that the transport properties of this model can be drastically changed by an interaction~\cite{KaneFisher-PRL1992, KaneFisher-PRB1992, Loss1992, Furusaki-Nagaosa1993, Gogolin1994, Meden2003, Dias2006}.
It should be clarified how the pathological behaviors of the Kohn--Drude weights are modified by the interaction.

It is also important to further develop the general theory for adiabatic transport. Our discussion in Sec.~\ref{general} suggests that  the defect energy $c_0(\theta)$ vanishes and that $c_{-1}(\theta)$ is a quadratic function of $\theta$, i.e., $d_n(\theta)=0$ ($n\geq3$) in any systems with a U(1) symmetry and a lattice translation symmetry, regardless of whether the low-energy effective theory of the system is TLL or not. In particular, $d_n(\theta)=0$ ($n\geq3$) is the condition for the linear Drude weight to be $\theta$-independent in the large-$L$ limit. These statements may be rationalized by the following argument. Let us imagine dividing the system into $M$ subsystems ($M\geq2$) in such a way that each part has the length $L_i\gg1$ and $\sum_{i=1}^ML_i=L$. We decompose $\theta$ correspondingly into $\theta_i\equiv\theta L_i/L$ ($i=1,\cdots,M$). We expect that the ground state energy of the $i$-th part is given by $E_0^{L_i,\theta_i}=\sum_{p=+1,0,-1,\cdots}c_p(\theta_i)L_i^p$ and that the total ground state energy satisfies the additivity $\sum_{i=1}^ME_0^{L_i,\theta_i}=E_0^{L,\theta}+O(L^{-1})$. This is possible only when $c_{0}(\theta)$, which should be independent of $\theta$, vanishes. Furthermore, we demand that the leading $L^{-1}$ correction is $\theta$-independent, i.e., $\sum_{i=1}^M[E_0^{L_i,\theta_i}-E_0^{L_i,0}]=[E_0^{L,\theta}-E_0^{L,0}]+o(L^{-1})$, which suggests that $c_{-1}(\theta)-c_{-1}(0)$ is proportional to $\theta^2$. We leave more rigorous proof of these conjectures to future work.

While we focused on the zero temperature limit in this work (except for the inequality Eq.~\eqref{eq:Kohn_bulk_fsum}), the distinction between the
Kohn--Drude and bulk Drude weights would be also important at finite temperatures.
Exploration of the problem at finite temperatures is also left for future investigations.

\begin{acknowledgments}
We thank Yohei Fuji, Hideaki Obuse, Yuhi Tanikawa, Hosho Katsura, Wonjun Lee, and Gil-Young Cho for valuable discussions. The work of K. T. was supported by the U.S. Department of Energy (DOE), Office of Science, Basic Energy Sciences (BES), under Contract No. AC02-05CH11231 within the Ultrafast Materials Science Program (KC2203), by JSPS KAKENHI Grant No.~JP22K20350, and by JST PRESTO Grant No.~JPMJPR2256. K. T. also thanks JSPS for support from Overseas Research Fellowship. The work of M. O. was supported in part by MEXT/JSPS KAKENHI Grant Nos.~JP17H06462 and JP19H01808, JST CREST Grant No. JPMJCR19T2. The work of H. W. was supported by JSPS KAKENHI Grant No.~JP20H01825 and JP21H01789, and by JST PRESTO Grant No.~JPMJPR18LA. 
\end{acknowledgments}


\appendix
\section{Expressions for tight-binding model}
\label{TBexpress}
Here we derive formulas for noninteracting fermions. The quadratic Hamiltonian $\hat{H}^{L,\theta} = \sum_{x,y=-L/2+1}^{L/2} \hat{c}^\dagger_x H_{xy}^{L,\theta} \hat{c}_y$ can be diagonalized as $\hat{H}^{L,\theta} = \sum_{n} \epsilon_n^{L,\theta} \hat{\gamma}^\dagger_n \hat{\gamma}_n$, where $\hat{\gamma}_n^\dagger$($\hat{\gamma}_n$) is creation (annihilation) operator for the $n$-th energy level $\epsilon_n^{L,\theta}$.  We arrange the index $n$ in such a way that $n<0$ ($n>0$) corresponds to occupied (unoccupied) states in the ground state.  We write $\epsilon_{mn}^{L,\theta}\equiv \epsilon_m^{L,\theta}-\epsilon_n^{L,\theta}$.

We expand operators in Eqs.~\eqref{eqj}--\eqref{eqt} with $\hat{\gamma}_{n}$ as 
\begin{align}
&\hat{j}^{L,\theta} =\frac{1}{L}\sum_{m,n} \hat{\gamma}^\dagger_m J_{mn}^{L,\theta} \hat{\gamma}_n, \label{eq:kmn_def} \\
&\hat{k}^{L,\theta} =\frac{1}{L}\sum_{m,n} \hat{\gamma}^\dagger_m K_{mn}^{L,\theta} \hat{\gamma}_n,\\
&\hat{t}^{L,\theta} =\frac{1}{L}\sum_{m,n} \hat{\gamma}^\dagger_m T_{mn}^{L,\theta} \hat{\gamma}_n. \label{eq:tmn_def}
\end{align}

Using these matrix elements, we obtain expressions for linear response functions:
\begin{align}
    &\phi^{L,\theta}_{(1),0}(\omega)=\frac{1}{L}\sum_{n < 0} K_{nn}^{L,\theta},  \label{eq:linear_opt_free_1} \\
    &\phi^{L,\theta}_{(1),1}(\omega)= \frac{1}{L}\sum_{\substack{n < 0,\\ m > 0}} |J_{mn}^{L,\theta}|^2 
    \left( \frac{1}{\omega-\epsilon_{mn}^{L,\theta} + i \eta}-\frac{1}{\omega+\epsilon_{mn}^{L,\theta} + i \eta} \right),\label{eq:linear_opt_free_2} \\
    &\tilde{\mathcal{D}}_{(1)}^{L,\theta} =\frac{1}{L} \sum_{n < 0} K_{nn}^{L,\theta}  - \frac{2}{L}\sum_{\substack{n < 0,\\ m > 0}} \frac{|J_{mn}^{L,\theta}|^2}{\epsilon_{mn}^{L,\theta}}, \label{eq:DKohn1_free} \\
    &\mathrm{Re}[\sigma_{(1)\mathrm{reg}}^{L,\theta}(\omega)]=\frac{\pi}{L}  \sum_{\substack{n < 0,\\ m > 0}} \frac{|J_{mn}^{L,\theta}|^2}{\epsilon_{mn}^{L,\theta}}\delta(|\omega|-|\epsilon_{mn}^{L,\theta}|),\\
  &  I_{(1)}^{L,\theta}(\Omega) = \tilde{\mathcal{D}}_{(1)}^{L,\theta} +\frac{2}{L}\sum_{\substack{n < 0, m > 0,\\
    \epsilon_{mn}^{L,\theta}<\Omega }} \frac{|J_{mn}^{L,\theta}|^2}{\epsilon_{mn}^{L,\theta}}. \label{eq:I1_free}
\end{align}

Similarly, second-order response functions are
\begin{align}
\phi_{(2),0}^{L,\theta}&=\frac{1}{L}\sum_{n < 0} T^{L,\theta}_{nn}
\end{align}
\begin{widetext}
\begin{align}
\phi_{(2),1}^{L,\theta}(\omega_1, \omega_2)&=\frac{1}{L}\sum_{\substack{n<0, \\m>0}} \left[ \mathcal{A}_{mn}^{L,\theta} \left\{ \frac{1}{\omega_1+\omega_2- \epsilon^{L,\theta}_{mn}+2i\eta}
-\frac{1}{\omega_1 - \epsilon^{L,\theta}_{mn}+i\eta}
-\frac{1}{\omega_2 - \epsilon^{L,\theta}_{mn}+i\eta}\right\} - (n \leftrightarrow m) \right],\\
\phi_{(2),2}^{L,\theta}(\omega_1, \omega_2)&=\frac{1}{L}\sum_{\substack{n, n^\prime<0,\\m, m^\prime>0}} \mathcal{B}_{m m^\prime n n^\prime}^{L,\theta} \left[
\left\{
\frac{1}{(\omega_1+\omega_2 - \epsilon_{mn}^{L,\theta} + 2i\eta) (\omega_1 - \epsilon_{m^\prime n^\prime}^{L,\theta} + i\eta)}  +\frac{1}{(\omega_1+\omega_2 - \epsilon_{mn}^{L,\theta} + 2i\eta) (\omega_2 - \epsilon_{m^\prime n^\prime}^{L,\theta} + i\eta)}\right. \right. \nonumber\nonumber\\
&\left. \left.\qquad \qquad \qquad \qquad \qquad -  \frac{1}{(\omega_1+  \epsilon_{mn}^{L,\theta} + i\eta) (\omega_2 - \epsilon_{m^\prime n^\prime}^{L,\theta} + i\eta)}\right\} + (m \leftrightarrow n^\prime, n \leftrightarrow m^\prime) \right],
\end{align}
\begin{align}
I_{(2)}^{L,\theta}(\Omega)
&=\tilde{\mathcal{D}}_{(2)}^{L,\theta} 
+ \frac{2}{L}\sum_{\substack{n<0,\\m>0}} \frac{\mathrm{Re}[\mathcal{A}_{mn}^{L,\theta}]}{\epsilon^{L,\theta}_{mn}}\left(g_1(\tfrac{\epsilon^{L,\theta}_{mn}}{\Omega})+2f_1(\tfrac{\epsilon^{L,\theta}_{mn}}{\Omega})\right)  
-\frac{2}{L}\sum_{\substack{n<0,\\m>0}} \frac{\mathrm{Im}[\mathcal{A}_{mn}^{L,\theta}]}{\epsilon^{L,\theta}_{mn}}\left(g_2(\tfrac{\epsilon^{L,\theta}_{mn}}{\Omega})-2f_2(\tfrac{\epsilon^{L,\theta}_{mn}}{\Omega})\right)\notag\\
&\quad-\frac{2}{L}\sum_{\substack{n,n'<0,\\m,m'>0}}\frac{\mathrm{Re}[\mathcal{B}_{m m^\prime n n^\prime}^{L,\theta}]}{\epsilon^{L,\theta}_{mn}\epsilon^{L,\theta}_{m'n'}}\left[
    2h_1(\tfrac{\epsilon^{L,\theta}_{m'n'}}{\Omega},\tfrac{\epsilon^{L,\theta}_{mn}}{\Omega})
    +f_1(\tfrac{\epsilon^{L,\theta}_{m'n'}}{\Omega})+f_1(\tfrac{\epsilon^{L,\theta}_{mn}}{\Omega})
    -f_1(\tfrac{\epsilon^{L,\theta}_{mn}}{\Omega})f_1(\tfrac{\epsilon^{L,\theta}_{m'n'}}{\Omega})
    -f_2(\tfrac{\epsilon^{L,\theta}_{mn}}{\Omega})f_2(\tfrac{\epsilon^{L,\theta}_{m'n'}}{\Omega})\right]
\notag\\
&\quad+\frac{2}{L}\sum_{\substack{n,n'<0,\\m,m'>0}}\frac{\mathrm{Im}[\mathcal{B}_{m m^\prime n n^\prime}^{L,\theta}]}{\epsilon^{L,\theta}_{mn}\epsilon^{L,\theta}_{m'n'}}\left[
    2h_2(\tfrac{\epsilon^{L,\theta}_{m'n'}}{\Omega},\tfrac{\epsilon^{L,\theta}_{mn}}{\Omega})
    +f_2(\tfrac{\epsilon^{L,\theta}_{m'n'}}{\Omega})
    -f_2(\tfrac{\epsilon^{L,\theta}_{mn}}{\Omega})
    -f_1(\tfrac{\epsilon^{L,\theta}_{mn}}{\Omega})f_2(\tfrac{\epsilon^{L,\theta}_{m'n'}}{\Omega})+f_1(\tfrac{\epsilon^{L,\theta}_{m'n'}}{\Omega})f_2(\tfrac{\epsilon^{L,\theta}_{mn}}{\Omega})\right], \label{eq:I2_free}
\end{align}
\begin{align}
\tilde{\mathcal{D}}_{(2)}^{L,\theta} =  \frac{1}{L}\sum_{n < 0} T^{L,\theta}_{nn} -\frac{6}{L}\sum_{\substack{n < 0, \\ m > 0}} \frac{\mathrm{Re}[\mathcal{A}_{mn}^{L,\theta}]}{\epsilon^{L,\theta}_{mn}}+\frac{6}{L}\sum_{\substack{n,n'<0,\\m,m'>0}} \frac{\mathcal{B}_{m m^\prime n n^\prime}^{L,\theta}}{\epsilon^{L,\theta}_{mn}\epsilon^{L,\theta}_{m'n'}}.\label{eq:DKohn2_free}
\end{align}
\end{widetext}

\section{Relation to  boundary conformal field theory}
\label{Sec:BCFT}
Here we mention the relation between our results and the boundary CFT. In fact, the first term of $c_{-1}(\theta)$ in Eq.~\eqref{cm} which depends on $T_F$ can be interpreted in terms of the boundary CFT. 

The defect in the tight-binding model we consider corresponds to the barrier in a TLL studied in Refs.~\cite{KaneFisher-PRL1992,KaneFisher-PRB1992} at the free fermion point $K=1$.
The system studied in this paper is a finite ring of circumference $L$ with a single defect.
Such a system can be mapped to a problem of boundary CFT by a folding trick~\cite{WongAffleck,IsingDefect-PRL,Yjunction-JSTAT}: after the folding, we have a 
two-component TLL of length $l=L/2$ with two boundaries: one corresponding to the defect and the other corresponding to no defect.
Although the problem at this point is a two-component TLL (of central charge 2) with boundaries, we can decompose the two-component TLL into
even and odd combinations of the original fields $\phi(x) \pm \phi(-x)$.
It can be seen that the odd component does not ``feel'' the defect and is always subject to the same boundary condition.
Thus the problem is effectively reduced to the single-component TLL (of the even field) with boundaries, although care should be taken about ``gluing condition''~\cite{WongAffleck,Yjunction-JSTAT,Oshikawa2010BCFT} in reconstructing the spectrum of the original model.
In the discussion of the universal part of the ground-state energy, which is the focus of the present paper, however, the gluing condition is not important and we can simply study the single-component TLL with boundaries.
For generic values of the Luttinger parameter $K$, the barrier is either a relevant or irrelevant perturbation (in the renormalization group sense), so that the defect is renormalized into an infinitely strong barrier which completely reflects the current, or a vanishing barrier which transmits the current perfectly.
In terms of the (even component of) ``phase field'' of the TLL, the infinitely strong barrier corresponds to the Neumann boundary condition, while the vanishing barrier corresponds to the Dirichlet boundary condition.
The free fermion case $K=1$ corresponds to the boundary between the two phases.
Here the barrier is an exactly marginal perturbation, so that there is a continuous family of the boundary conditions interpolating the vanishing barrier and the infinitely strong barrier. This exactly corresponds to the $S$-matrix of the defect for the incoming free fermions.

In fact, the continuous family of the boundary conditions at $K=1$ was studied in terms of free bosons in Refs.~\cite{CallanKlebanov,CKLM1994} and in terms of free fermions in Ref.~\cite{PolchinskiThorlacius}.
Here we show that our result agrees with theirs.
In Refs.~\cite{CallanKlebanov,CKLM1994}, the continuous family of the boundary conditions at $K=1$ is formulated in terms of the emergent SU(2) symmetry.
This SU(2) degree of freedom indeed corresponds to the $S$-matrix of free fermions.
In Ref.~\cite{PolchinskiThorlacius}, the $S$-matrix was parametrized in terms of the effective barrier parameter $g$ (complex number).
For simplicity here we focus on the case $g$ is real, for which the $S$-matrix is given as $S \sim \exp{(i \pi g \sigma_1)}$ and thus $T^\pm_F = T_F = \cos{(\pi g)}$.
The phase shift in those papers, for example in Eq.~(13) of Ref.~\cite{PolchinskiThorlacius}, then simply reads as $\Delta = \pi |g| = \arccos{T_F}$.

The partition function for the boundary condition corresponding to the barrier strength $g$ on one side and the Neumann boundary condition (corresponding to the infinite barrier) on the other side (in the even fermion number sector) at the inverse temperature $\beta$ is~\cite{PolchinskiThorlacius}
\begin{equation}
    Z_{BN}(q) \sim \frac{q^{-1/24}}{\prod_{m=1}^\infty (1-q^m)} \sum_{k=0}^\infty q^{\frac{1}{2} \left[ k + \frac{1}{2} - (-1)^k \frac{\Delta}{\pi} \right]}, 
\end{equation}
where $q \equiv e^{-\pi \beta/l}$.
In our setup, we just consider a single barrier, so that the other side obeys the Dirichlet boundary condition after the folding.
In this case, the partition function is rather given by
\begin{equation}
    Z_{BD}(q) \sim \frac{q^{-1/24}}{\prod_{m=1}^\infty (1-q^m)} \sum_{k=0}^\infty q^{\frac{1}{2} \left[ k +  (-1)^k \frac{\Delta}{\pi} \right]} .
\end{equation}
We can read off the ground-state energy (relative to $g=0$ case) as
\begin{equation}
    \frac{\pi}{l} \frac{1}{2} \left( \frac{\Delta}{\pi} \right)^2 = \frac{1}{L} \frac{\Delta^2}{\pi} . 
\end{equation}
Because $\Delta = \arccos{T_F}$, this agrees with our Eq.~\eqref{cm} with $\delta_F=\theta=\varphi_F=0$ and $v_F=1$.

\section{Perturbative justification of the main result}
\label{perturbation}
Some key aspects of $c_{-1}(\theta)$ derived in Sec.~\ref{GSE} of the main text can be readily understood by treating $\hat{V}$ as perturbation. Here we use the single impurity potential~\eqref{impuritypotential} as an example. As noted before, $r_F\equiv R_F/T_F=|w|/v_F$ and $\delta_F=0$ in this case. The matrix element $\langle k_n|\hat{V}| k_n\rangle=w/L$ is independent of $n$ and $m$ and is inversely proportional to the system size. 

The first-order correction to the ground-state energy is
\begin{equation}
\sum_{n:\mathrm{occ}.}\langle k_n|\hat{V}| k_n\rangle=w\frac{N_{\mathrm{el}}}{L}
\end{equation}
This is a part of the defect energy $c_0(\theta)$.

\begin{figure*}[t]
    \begin{center}
    \includegraphics[width=14cm]{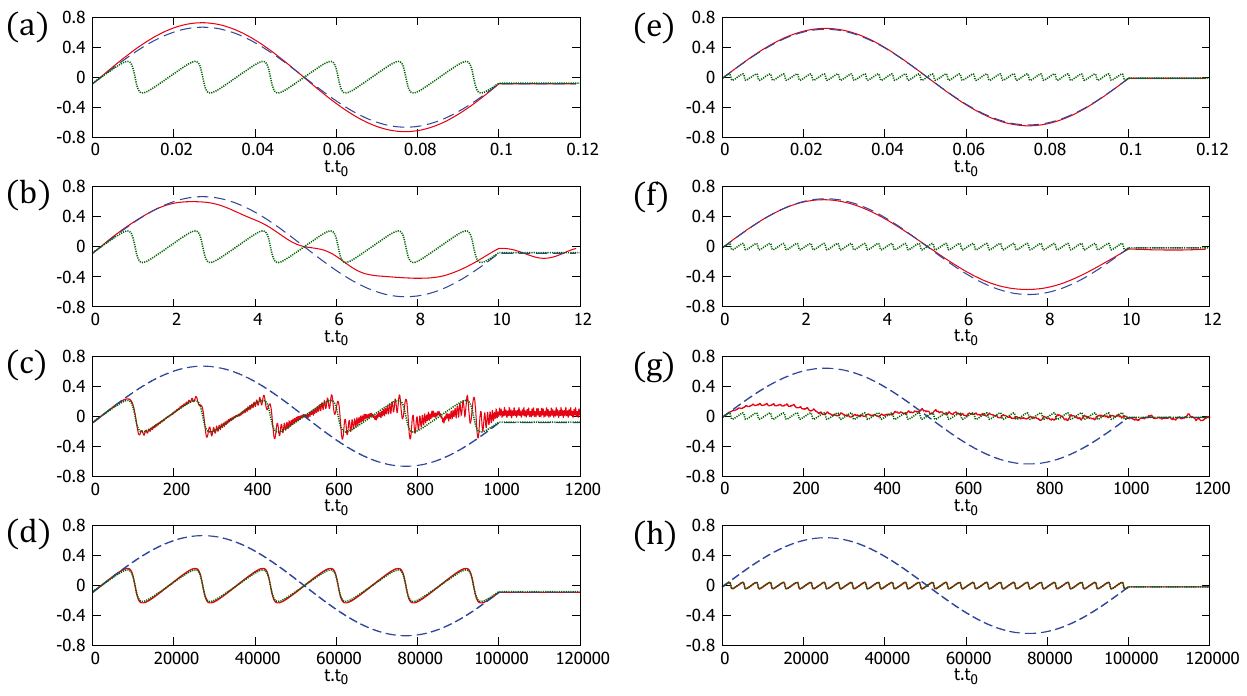}
    \caption{\label{fig3}
    (a--d) [(e--f)] Real-time evolution of the current density $j(t)$ (red curve) for $L=6$ [$L=30$] driven by the time-dependent Hamiltonian with a bond disorder~\eqref{H(t)_v}. The ramp time is set to $T=10^{-1}/t_0, 10/t_0, 10^3/t_0$ and $10^5/t_0$ in (a), (b), (c) and (d) [(e), (f), (g) and (h)], respectively. The total flux $A_0$ is $2\pi$. The parameters for the disorder are set to $v=1.5t_0$ and $\delta=\pi/4$. The blue and green dashed curves represent the adiabatic current density for $v=0$ and $v \neq 0$.
    }
    \end{center}
    \end{figure*}

The second-order correction is given by
\begin{align}
&-\sum_{n:\mathrm{occ.}}\sum_{m:\mathrm{unocc.}}\frac{\langle k_n|\hat{V}| k_m\rangle\langle k_m|\hat{V}|k_n\rangle}{\epsilon_{k_m}-\epsilon_{k_n}}\notag\\
&=-\frac{v_F^2r_F^2}{L^2}\sum_{n:\mathrm{occ.}}\sum_{m:\mathrm{unocc.}}\frac{1}{\epsilon_{k_m}-\epsilon_{k_n}}
\label{E2}
\end{align}
To extract the most singular contributions from adjacent of the Fermi points $k=\pm k_F$, we linearize the dispersion as $\epsilon_{k}=\pm v_F(k \mp k_F+\theta/L)$.  The summation in Eq.~\eqref{E2} is dominated by the scattering between $k=k_F$ and $-k_F$:
\begin{align}
&-\frac{v_Fr_F^2}{L^2}\sum_{k_n>-k_F}\sum_{k_m>k_F}\frac{1}{k_m+k_n+2\theta/L}\notag\\
&\quad\quad+\frac{v_Fr_F^2}{L^2}\sum_{k_n<k_F}\sum_{k_m<-k_F}\frac{1}{k_m+k_n+2\theta/L}\notag\\
&=-\frac{v_Fr_F^2}{2\pi L}\sum_{n,m=0}^\infty\frac{2(m+n+1)}{(m+n+1)^2-(\theta/\pi)^2}\notag\\
&=\frac{v_Fr_F^2}{2\pi L}\theta\cot \theta+\dots.
\end{align}
In the last step, we regularize the summation by subtracting the $\theta$-independent term.  This gives the correct $\theta$-dependence of $c_{-1}(\theta)$ in Eq.~\eqref{cm} up to $r_F^2$, and Eqs.~\eqref{d1}--\eqref{d3} can be fully reproduced up to this order of $r_F$.

The above perturbation theory fails near $\theta=\pi$ because of the degeneracy of the $N_{\mathrm{el}}$-th level and the $(N_{\mathrm{el}}+1)$-th level ($n=\ell$ and $n=-\ell-1$) of $\hat{H}_0^{L,\theta}$.
Focusing only on these two levels, we find
\begin{equation}
\begin{pmatrix}
\epsilon_{k_{\ell}}&0\\
0&\epsilon_{k_{-\ell-1}}
\end{pmatrix}
+\frac{w}{L}
\begin{pmatrix}
1&1\\
1&1
\end{pmatrix}.
\end{equation}
The smaller eigenvalue of this matrix is
\begin{align}
&\frac{w}{L}+\varepsilon_F\cos\left(\frac{\theta-\pi}{L}\right)-\sqrt{v_F^2\sin^2\left(\frac{\theta-\pi}{L}\right)+\frac{w^2}{L^2}}, 
\end{align}
implying that $c_{-1}(\theta)$ contains
\begin{align}
&-v_F\sqrt{(\theta-\pi)^2+r_F^2}\notag\\
&=-v_F\sum_{N=0}^\infty \frac{\Gamma(\tfrac{3}{2})}{N!\Gamma(\tfrac{3}{2}-N)}(\theta-\pi)^{2N}r_F^{1-2N}.
\end{align}
Hence, the most singular term in $d_{2N}(\theta=\pi)$ ($N\geq 1$) in the $r_F=+0$ limit is given by 
\begin{align}
-v_F\frac{(2N)!\Gamma(\tfrac{3}{2})}{N!\Gamma(\tfrac{3}{2}-N)}\frac{1}{r_F^{2N-1}}.
\end{align}
This reproduces our results in Eqs.~\eqref{s2} and \eqref{s4}.

\section{Finiteness of matrix elements}
\label{App:finiteness}
In this appendix, we prove that the matrix elements $J_{mn}^{L,\theta}$, $K_{mn}^{L,\theta}$, and $T_{mn}^{L,\theta}$, defined in Eqs~\eqref{eq:kmn_def}-\eqref{eq:tmn_def}, are $O(1)$ quantity in general.

Let us consider a system of noninteracting fermions defined on a lattice $\Lambda$. The Hamiltonian is given by
\begin{equation}
\hat{H}\equiv\sum_{x,y\in\Lambda}\hat{c}_x^\dagger H_{xy}\hat{c}_y=\hat{\bm{c}}^\dagger H\hat{\bm{c}}.
\end{equation}
In the last expression, we regarded $\hat{c}_x$ as a component of a columnar vector $\hat{\bm{c}}$ and $H_{xy}$ as matrix elements of a Hermitian matrix $H$. Diagonalizing $H$ by a unitary matrix $U$, we find
\begin{equation}
\hat{H}=\sum_n\epsilon_n\hat{\gamma}_n^\dagger \hat{\gamma}_n,
\end{equation}
where $\gamma_n^\dagger\equiv\sum_{x\in\Lambda}\hat{\bm{c}}^\dagger\bm{u}_{n}$ and $\bm{u}_{n}$ is the $n$-th columnar vector of the unitary matrix $U$.

Now we consider an operator $\hat{O}$ that takes the form
\begin{align}
\hat{O}&\equiv\hat{\bm{c}}^\dagger O\hat{\bm{c}}=\sum_{x,y\in\Lambda}\hat{c}_x^\dagger O_{xy}\hat{c}_y.
\end{align}
We assume that $\hat{O}$ is finite ranged; i.e., $O_{xy}$ vanishes when $|x-y|$ is larger than the range $R$. We rewrite $\hat{O}$ in the basis of $\hat{\gamma}_n$:
\begin{align}
\hat{O}&=\sum_{n,m}\hat{\gamma}_n^\dagger (U^\dagger OU)_{nm}\hat{\gamma}_m=\sum_{n,m}\hat{\gamma}_n^\dagger O_{nm}\hat{\gamma}_m.
\end{align}
In the following, we show that matrix elements $O_{nm}\equiv\bm{u}_n^\dagger O\bm{u}_{m}$ can be bounded by a system-size independent constant. 

Writing $M\equiv \max_{x,y\in\Lambda}|O_{xy}|$, we have
\begin{align}
|O_{nm}|
&\leq \sum_{x,y\in\Lambda}|\bm{u}_n|_x |O_{xy}||\bm{u}_{m}|_y\notag\\
&\leq M  \sum_{x,y\in\Lambda}|\bm{u}_n|_x D_{xy}|\bm{u}_{m}|_y=M\bm{v}_n^TD\bm{v}_m
\end{align}
Here, we introduced normalized vectors $\bm{v}_n$ by $(\bm{v}_n)_x=|\bm{u}_n|_x$ and a real symmetric matrix $D$ by 
\begin{align}
D_{xy}=\begin{cases}
1& |x-y|\leq R\\
0& |x-y|>R.
\end{cases}
\end{align}
The eigenvalues of $D$ can be easily found, which are given by $\epsilon_k=1+\sum_{1\leq z\leq R}2\cos kz$ with $k=2\pi j/L$. We have
\begin{align}
|\epsilon_k|\leq 1+2R.
\end{align}
Therefore,
\begin{align}
|O_{nm}|\leq M\bm{v}_n^TD\bm{v}_m\leq M(1+2R).
\end{align}
Generalization to multi-band cases is straightforward.

\section{Real-time simulation for a bond disorder}
\label{AppC}
Our main claim in this paper does not depend on the detail of the defect. This is because the universal dependence of the ground state energy on the twist angle is fully characterized by the transmission coefficient of the defect scattering as shown in Sec.~\ref{sec:twist}. To support this point from the numerical calculation, here we examine the bond disorder in~\eqref{bonddisorder} as our second example.

We apply a static electric field to the tight-binding model with the bond disorder. The time-dependent Hamiltonian is given by
\begin{align}
\hat{H}^{L,\theta(t)}&=-t_0\sum_{x=-L/2+1}^{L/2}(\hat{c}_{x+1}^\dagger e^{-i\theta(t)/L}\hat{c}_x+\mathrm{h.c.}) \nonumber\\
&\qquad+ \{-(ve^{i\delta}-t_0)\hat{c}_{1}^\dagger e^{-i\theta(t)/L}\hat{c}_{0}+\mathrm{h.c.}\},
\label{H(t)_v}
\end{align}
where the systems size 
$L=4\ell+2$ and the number of electrons $N_{\mathrm{el}}=L/2$. The time dependence of the flux $\theta(t)$ is given by Eq.~\eqref{theta(t)}. The time-evolution protocol and the calculation method are the same as in Sec.~\ref{sec:realtime}.

We calculate the current density $j^L(t)$ defined by Eq.~\eqref{current} and the results are shown in Fig.~\ref{fig3}. The qualitative behavior is the same as the potential disorder case shown in~Fig.~\ref{fig2}.  This supports the validity of our analytical results in Sec.~\ref{sec:twist}. Also, this result suggests that the current response is weakened by the defect while the defect induces the divergence of nonlinear Drude weights. A slight difference from the potential disorder is the time-reversal symmetry breaking effect, i.e., $\delta \neq 0$. This makes the persistent current nonvanishing even without an electric field.

\bibliography{ref.bib}

\clearpage

\end{document}